\documentclass{aa}  

\usepackage{graphicx}
\usepackage{txfonts}
\usepackage{orcidlink}
\begin{document} 
        
        \title{KiDS-1000: Cross-correlation with Planck cosmic microwave background lensing and intrinsic alignment removal with self-calibration}

        
        \author{
                Ji Yao\inst{1,2,3}\thanks{\email{ji.yao@shao.ac.cn}}\orcidlink{0000-0002-7336-2796},
                Huanyuan Shan\inst{1}\thanks{\email{hyshan@shao.ac.cn}}\orcidlink{0000-0001-8534-837X},
                Pengjie Zhang\inst{2,3,4}\thanks{\email{zhangpj@sjtu.edu.cn}},
                Xiangkun Liu\inst{5},
                Catherine Heymans\inst{6,7},
                Benjamin Joachimi\inst{8},
                Marika Asgari\inst{9},
                Maciej Bilicki\inst{10},
                Hendrik Hildebrandt\inst{6},
                Konrad Kuĳken\inst{11},
                Tilman Tröster\inst{6},
                Jan Luca van den Busch\inst{8,12},
                Angus Wright\inst{7},
                \and
                Ziang Yan\inst{7}
        }
        
        \institute{
                Shanghai Astronomical Observatory (SHAO), Nandan Road 80, Shanghai 200030, China
                \and
                Department of Astronomy, School of Physics and Astronomy, Shanghai Jiao Tong University, Shanghai, 200240, China
                \and
                Shanghai Key Laboratory for Particle Physics and Cosmology, Shanghai 200240, China
                \and
                Tsung-Dao Lee Institute, Shanghai, 200240, China
                \and
                South-Western Institute for Astronomy Research, Yunnan University, Kunming, 650500, China
                \and
                Institute for Astronomy, University of Edinburgh, Royal Observatory, Blackford Hill, Edinburgh, EH9 3HJ, UK
                \and
                Ruhr-Universität Bochum, Astronomisches Institut, German Centre for Cosmological Lensing (GCCL), Universitätsstr. 150, 44801, Bochum, Germany
                \and
                Department of Physics and Astronomy, University College London, Gower Street, London WC1E 6BT, UK
                \and
                E.A Milne Centre, University of Hull, Cottingham Road, Hull, HU6 7RX, United Kingdom
                \and
                Center for Theoretical Physics, Polish Academy of Sciences, al. Lotników 32/46, 02-668 Warsaw, Poland
                \and
                Leiden Observatory, Leiden University, P.O. Box 9513, 2300 RA Leiden, the Netherlands
                \and
                Argelander-Institut für Astronomie, Universität Bonn, Auf dem Hügel 71, 53121 Bonn, Germany
        }
        
        \date{Received January 30, 2023; accepted ?}
        
        
\abstract
{Galaxy shear and cosmic microwave background (CMB) lensing convergence cross-correlations contain additional information on cosmology with respect to auto-correlations. While remaining immune to certain systemic effects, these cross-correlations are nonetheless affected by the galaxy's intrinsic alignments (IA). These effects may, in fact, be responsible for the reported low lensing amplitude of the galaxy shear $\times$ CMB convergence cross-correlations, compared to the standard Planck $\Lambda$CDM (cosmological constant and cold dark matter) cosmology predictions.}
{In this work, we investigate how IA affects the Kilo-Degree Survey (KiDS) galaxy lensing shear and Planck CMB lensing convergence cross-correlation and we compare it to previous treatments, both with and without IA taken into consideration.}
{We compared the marginalization over IA parameters and the IA self-calibration (SC) method (with additional observables defined only from the source galaxies) to demonstrate that SC can efficiently break the degeneracy between the CMB lensing amplitude, $A_{\rm lens}$, and the IA amplitude, $A_{\rm IA}$. We further investigated how different systematics affect the resulting $A_{\rm IA}$ and $A_{\rm lens}$ and we validated our results with the MICE2 simulation.}
{We find that by including the SC method to constrain IA, the information loss due to the degeneracy between CMB lensing and IA is strongly reduced. The best-fit values are $A_{\rm lens}=0.84^{+0.22}_{-0.22}$ and $A_{\rm IA}=0.60^{+1.03}_{-1.03}$, while different angular scale cuts can affect $A_{\rm lens}$ by $\sim10\%$. We show that an appropriate treatment of the boost factor, cosmic magnification, and photometric redshift modeling is important for obtaining the correct IA and cosmological results.}
{}

\keywords{cosmology --
        weak lensing --
        CMB lensing --
        intrinsic alignment --
        self-calibration
}
\titlerunning{KiDS shear $\times$ Planck lensing  and IA removal}
\authorrunning{Yao et al 2022} 
\maketitle

%

\section{Introduction}

Weak lensing due to the distortion of light by gravity is a powerful
probe of the underlying matter distribution and the encoded information on cosmological physics such as dark matter, dark energy, and the nature
of gravity \citep{Refregier2003,Mandelbaum2018}. The auto-correlation statistics have been widely used in such analyses, both for galaxy lensing shear, for instance, ``cosmic shear'' \citep{Hildebrandt2016,HSC_Hamana2020,HSC_Hikage2019,Asgari2021,DESY3model,DESY3data},
as well as cosmic microwave background (CMB) lensing convergence \citep{Planck2018lensing,Omori2017}. Furthermore, cross-correlations between galaxy shear and CMB lensing have been measured extensively \citep{Hand2015,Chisari2015,Liu2015,Kirk2016,Harnois-Deraps2016,Singh2017b,Harnois-Deraps2017,Omori2019,Namikawa2019,Marques2020,Robertson2021}.
Cross-correlation statistics contain highly complementary information with respect to auto-correlations, both for cosmology and the cross-checking of systematics. They partly reveal the hidden redshift information in CMB
lensing and are more sensitive to structure growth at redshifts between the epochs probed by galaxy shear and CMB lensing. Cross-correlations are also immune to additive errors in shear measurement and provide an external diagnosis of multiplicative errors \citep{Schaan2017}.

Most existing cross-correlation measurements have found a lower CMB lensing amplitude than the prediction of their assumed $\Lambda$CDM cosmology \citep{Hand2015,Liu2015,Kirk2016,Harnois-Deraps2016,Harnois-Deraps2017,Singh2017b,Marques2020,Robertson2021}. This ratio is normally referred as the CMB lensing amplitude, $A_{\rm lens}\sim 0.5$-$0.9$, although the deviation from unity
is only within $1$-$2\sigma$. The low lensing amplitude is
consistent across many combinations of data sets and analytical
methods, suggesting the existence of a common systematic
errors or a deviation from the best-fit {\it Planck} cosmology. This might be related to the tension between galaxy lensing surveys and {\it Planck} CMB observations \citep{Lin2017b,Chang2019,Heymans2021} with  {\it Planck's} internal inconsistencies \citep{Planck2018I,Planck2018VIcosmo}. In this paper, we focus on the galaxy
intrinsic alignment (IA), which can mimic weak lensing signals \citep{Croft2000,Catelan2001,Crittenden2001,Lee2002,Jing2002,Hirata2004,Heymans2004,BridleKing,Okumura2009,Joachimi2013,Kiessling2015,Blazek2015,Rong2015,Krause2016,Blazek2017,Troxel2018,Chisari2017,Xia2017,Samuroff2019,Yao2019,Samuroff2021,Yao2020}. Here, the CMB lensing convergence is expected to be anticorrelated with respect to the intrinsic ellipticities of the foreground galaxy field, resulting in a dilution of the overall cross-correlation signal \citep{Troxel2014,Chisari2015,Kirk2015,Omori2019,Robertson2021}.
Taking IA into account can alleviate the tension in $A_{\rm lens}$, at the expense of a significant loss of lensing constraining power because of the degeneracy between the lensing amplitude, $A_{\rm lens}$, and the IA amplitude, $A_{\rm IA}$. Therefore, a common compromise is to fix
both the IA model and its amplitude $A_{\rm IA}$
\citep{Kirk2016,Harnois-Deraps2017,Omori2019} or to assume a strong prior \citep{Robertson2021}.  

Since IA is already a major limiting factor in the current cross-correlation
analysis, its mitigation will be essential for upcoming
measurements with significantly smaller statistical errors.
We utilize the IA
self-calibration (SC) method \citep{Zhang2010,SC2008,Troxel2012b,Troxel2012,Yao2017,Yao2018}, a galaxy-galaxy lensing method that is based on a different weighting scheme, to mitigate the IA problem in the shear-convergence cross-correlation. It draws on the fact that the IA-galaxy correlation is insensitive to the redshift order, while it does indeed matter for lensing-galaxy correlations whether the lens is in front of the source or not.
Therefore, we can isolate IA by comparing extra observables, namely, the galaxy shear $\times$ number density cross-correlation with a different weighting of the redshift pairs. This measurement of IA is independent of a physical model of the IA and requires no data external to the shear data.  SC was first applied to KiDS450/KV450 \citep{Yao2019,Pedersen2020} and DECaLS DR3
\citep{Yao2020} and has enabled significant IA
detections. The detected IA signal can then be applied to remove IA in the lensing shear auto-correlation and shear-convergence cross-correlation. The IA information is obtained from a shear $\times$ number density cross-correlation within the same photometric redshift (photo-z) bin and (more importantly) with different weighting schemes on the photo-z ordering, which is usually not used for cosmological parameter constraints. We find that this removal of IA leads no almost no losses in terms of cosmological information.  

In a previous work \cite{Yao2020}, we demonstrated the importance and methodology of including certain types of systematics in the SC lensing-IA separation method, namely, galaxy bias, covariance between the separated lensing signal and IA signal, IA signal drop $Q^{\rm Ig}$ due to the photo-z selection, and scale dependency among the signal drops, $Q^{\rm Gg}$ and $Q^{\rm Ig}$. In this work, we further investigate other sources of systematics, including the boost factor \citep{Mandelbaum2005boostfactor}, photo-z modeling bias \citep{Yao2019}, and cosmic magnification \citep{Bartelmann1995,Bartelmann2001,Yang2017,Liu2021}. Interestingly, as the survey goes to higher redshift, the contamination to the SC method from magnification quickly increases to a non-negligible level. The cosmic magnification will change the observed galaxy number density due to the lensing-magnified flux and lensing-enlarged area, thereby biasing our SC analysis. We investigate the proper treatments for the above systematics together with the cosmological study.

This paper is organized as follows. In Sect.\,\ref{Section method}, we review the physics of galaxy shear $\times$ CMB convergence and how our SC method works to subtract the IA information. In Sect.\,\ref{Section data}, we introduce the KiDS-1000 and {\it Planck} data used in this work, and the MICE2 simulation \citep{vandenBusch2020,Fosalba2015} we use to validate how the SC method is affected by different systematics. We show the measurements of the observables in Sect.\,\ref{Section measurements}. The results and summary are shown in Sects.\,\ref{Section results} and \ref{Section summary}.

\section{Methods} \label{Section method}

We applied our self-calibration method to separate the intrinsic alignment and the lensing signals and show how the intrinsic alignment will bias the galaxy shear-CMB convergence correlation. In this section, we review the theory of lensing cross-correlation and the self-calibration method, with a modification to account for the contamination from cosmic magnification.

\subsection{Galaxy shear $\times$ CMB convergence} \label{Section shear X convergence}

The gravitational field can distort the shape of the background source galaxy image and introduce an extra shape that is tangentially aligned to the lens. This gravitational shear $\gamma^{\rm G}$ of the source galaxy contains integral information of the foreground overdensity along the line of sight \citep{Bartelmann2001}. Similarly, the photons from the CMB are deflected, and the lensing convergence, $\kappa,$ can be reconstructed from the CMB temperature and polarization observations \citep{Planck2018lensing}. By correlating these two quantities $\left <\gamma^{\rm G}\kappa\right >$, we probe the clustering of the underlying matter field $\left<\delta\delta\right>$. In harmonic space, while assuming flat space \citep{Omori2019,Marques2020}, we have:
\begin{equation}
        C^{\kappa^{\rm gal}\kappa^{\rm CMB}}(\ell)=\int_{0}^{\chi_{\rm CMB}}\frac{q^{\rm gal}(\chi)q^{\rm CMB}(\chi)}{\chi^2}P_\delta\left(k=\frac{\ell+1/2}{\chi},z\right)d\chi. \label{eq C^GK}
\end{equation}

Equation\,\ref{eq C^GK} is the galaxy-lensing CMB-lensing cross-angular power spectrum, which probes the matter power spectrum, $P_\delta(k,z)$, as well as the background geometry, $\chi(z),$ when precision allows. Here, $z$ is the redshift, $\chi$ is the comoving distance, $k$ is the wavenumber, $\ell$ is the angular mode, $q^{\rm gal}(\chi)$ and $q^{\rm CMB}(\chi)$ are the lensing efficiency functions for galaxy-lensing and CMB-lensing, with the analytical forms:
\begin{align}
        &q^{\rm gal} (\chi_{\rm l}) = \frac{3}{2}\Omega_{\rm m}\frac{H_0^2}{c^2}(1+z_{\rm l})
        \int_{\chi_l}^\infty
        n(\chi_{\rm s})\frac{(\chi_{\rm s}-\chi_{\rm l})\chi_{\rm l}}{\chi_{\rm s}}d\chi_{\rm s}, \label{eq q gal}\\
        &q^{\rm CMB} (\chi_{\rm l}) = \frac{3}{2}\Omega_{\rm m}\frac{H_0^2}{c^2}(1+z_{\rm l})
        \frac{(\chi_{\rm s}-\chi_{\rm l})\chi_{\rm l}}{\chi_{\rm s}}, \label{eq q CMB}
\end{align}
where $\chi_{\rm s}$ and $\chi_{\rm l}$ are the comoving distance to the source and lens, and $\chi_{\rm s}$ in Eq.\,\ref{eq q CMB} takes CMB as the source of light ($z\sim1100$). We note the spacial curvature $\Omega_k=0$ is assumed so that the comoving angular diameter distances in Eqs.\,\ref{eq q gal} and \ref{eq q CMB} are replaced with the comoving radial distances. Here, $n(\chi)$ gives the source galaxy distribution as a function of comoving distance and it is connected with the galaxy redshift distribution via $n(\chi)=n(z) dz/d\chi$. In this work, we only use one redshift bin due to the limit of the total S/N on the CMB lensing signal, while a tomographic example can be found in \cite{Harnois-Deraps2017}. In the future, with higher S/N (e.g., for CMB-S4\footnote{\url{https://cmb-s4.org/}} $\times$ LSST\footnote{Legacy Survey of Space and Time, Vera C. Rubin Observatory, \url{https://www.lsst.org/}}), tomography can be used to subtract more cosmological information.

The shear-convergence cross-correlation function measured in real space is given by the Hankel transformation:
\begin{equation}
        w^{\rm G\kappa}(\theta) = \frac{1}{2\pi}\int_{0}^{\infty}d\ell \ell C^{\kappa^{\rm gal}\kappa^{\rm CMB}}(\ell) J_2(\ell\theta) \label{eq w Hankel},
\end{equation}
where $J_2(x)$ is the Bessel function of the first kind and order 2. Here, ``G'' represents the gravitational lensing shear, $\gamma^{\rm G}$, that is to be separated from the intrinsic alignment, $\gamma^{\rm I}$, in the following section.

In addition, due to the current low S/N, we chose not to investigate the full cosmological constraints in this work. Instead, we performed a matched-filter fitting, with a lensing amplitude, $A_{\rm lens}$, that suits
$\hat{w}^{\rm G\kappa} = A_{\rm lens} {w}^{\rm G\kappa}$, where $\hat{w}^{\rm G\kappa}$ is the measured correlation function and ${w}^{\rm G\kappa}$ is the theoretical model.

\subsection{Intrinsic alignment of galaxies} \label{Section IA}

The observed galaxy shear estimator contains three components: gravitational shear, an intrinsic alignment term, and random noise, namely, $\hat{\gamma}=\gamma^{\rm G}+\gamma^{\rm I}+\gamma^{\rm N}$. Both the gravitational shear and the IA term are related to the underlying matter overdensity $\delta$ and are associated with the large-scale structure. This means that when we cross-correlate the galaxy shape and the CMB convergence, there will be contributions from both lensing and IA:
\begin{equation}
        \left < \hat{\gamma}\kappa \right >=\left <\gamma^{\rm G}\kappa \right >+\left <\gamma^{\rm I}\kappa \right >.
\end{equation}
Therefore, the IA part of the correlation will contaminate the measurement and lead to a bias in the lensing amplitude, $A_{\rm lens}$, or the cosmological parameters when assuming $\left<\hat{\gamma}\kappa\right>=\left<\gamma^{\rm G}\kappa\right>$.

The IA-convergence correlation function is linked to the IA-convergence power spectrum:
\begin{equation}
        C^{\rm I\kappa^{\rm CMB}} = \int_{0}^{\chi_{\rm CMB}}\frac{n(\chi)q^{\rm CMB}(\chi)}{\chi^2}P_{\delta,\gamma^{\rm I}}\left(k=\frac{\ell+1/2}{\chi},z\right)d\chi. \label{eq C^IK}
\end{equation}
Here, $P_{\delta,\gamma^{\rm I}}$ is the 3D matter-IA power
spectrum. The conventional method is to assume an IA model with some nuisance parameters, which will enter the fitting process. The most widely used IA model is the non-linear linear tidal alignment model \citep{Catelan2001,Hirata2004,BridleKing}, expressed as:
\begin{equation}
        P_{\delta,\gamma^{\rm I}}=-A_{\rm IA}(L,z)\frac{C_1\rho_{\rm m,0}}{D(z)}P_\delta(k;\chi), \label{eq IA P(k) model}
\end{equation}
which is proportional to the non-linear matter power spectrum $P_\delta$, suggesting that the IA is caused by the gravitational tidal field. Then, $A_{\rm IA}$ is the IA amplitude, which can be redshift($z$)- and luminosity($L$)- dependent \citep{Joachimi2011}. Its redshift evolution has been measured recently in simulations \citep{Chisari2016,Samuroff2021} and in observations with low significance \citep{Johnston2019,Yao2020,DESY3model,Tonegawa2022}. The other related quantities include: the mean matter density
of the universe at $z=0$, expressed as $\rho_{\rm m,0}=\rho_{\rm crit}\Omega_{\rm m,0}$; $C_1=5\times 10^{-14}(h^2M_{\rm sun}/{\rm
        Mpc}^{-3})$ the empirical amplitude taken from
\cite{Brown2002} and the normalized linear growth
factor $D(z)$. We note that the IA model in Eq.\,\ref{eq IA P(k) model} can be replaced by more complicated models as in \cite{Krause2016,Blazek2015,Blazek2017,Fortuna2021} for different samples \citep{Yao2020,Samuroff2021,Zjupa2020}. The self-calibration method can introduce new observables to constrain IA with additional constraining power, and in the future when the signal-to-noise (S/N) allows, it can be extended to constrain more complicated IA models.

\subsection{Self-calibration of intrinsic alignment} \label{Section SC}

The IA self-calibration (SC) method \citep{SC2008,Yao2017,Yao2018,Yao2019,Yao2020} uses the same galaxy sample as both the source and the lens, which is different from most galaxy-galaxy lensing studies. It introduces two observables: the shape-galaxy correlation in the same redshift bin, $w^{\rm \gamma g}$, and a similar correlation, $w^{\rm \gamma g}|_{\rm S}$, using the pairs where the photo-z of the source galaxy is lower than the photo-z of the lens galaxy, namely: 
\begin{equation}
        z^{\rm P}_\gamma<z^{\rm P}_{\rm g} \label{eq SC selection}
,\end{equation} which is denoted in this work as ``the SC selection.'' 

In this work, we extend our methodology to include the impact from cosmic magnification \citep{Bartelmann1995,Bartelmann2001,Yang2017,Liu2021}. Because of the existence of magnification, the intrinsic galaxy number density field, $\delta_g$, is affected by the foreground lensing convergence, $\kappa^{\rm gal}$, leading to a lensed galaxy overdensity:
\begin{equation}
        \delta^{\rm L}_{\rm g} = \delta_{\rm g} + g_{\rm mag}\kappa^{\rm gal}, \label{eq lensed overdensity}
\end{equation}
where the prefactor writes $g_{\rm mag}=2(\alpha-1)$ for a complete and flux-limited sample. It accounts for the increase in galaxy number density due to lensing-magnified flux ($\alpha=-d\ln N/d\ln F$, where $N(F)$ denotes the galaxy number $N$ that is brighter than the flux limit $F$) and the decrease of galaxy number density due to the lensing-area-enlargement (-2 in $g_{\rm mag}$). The observed shape-galaxy correlation is given by:
\begin{equation}
        \left <\hat{\gamma} \delta^{\rm L}_{\rm g} \right > = \left <(\gamma^{\rm G}+\gamma^{\rm I})(\delta_{\rm g}+g_{\rm mag}\kappa^{\rm gal})\right >.
\end{equation}

\begin{figure}\centering
        \includegraphics[width=1.0\columnwidth]{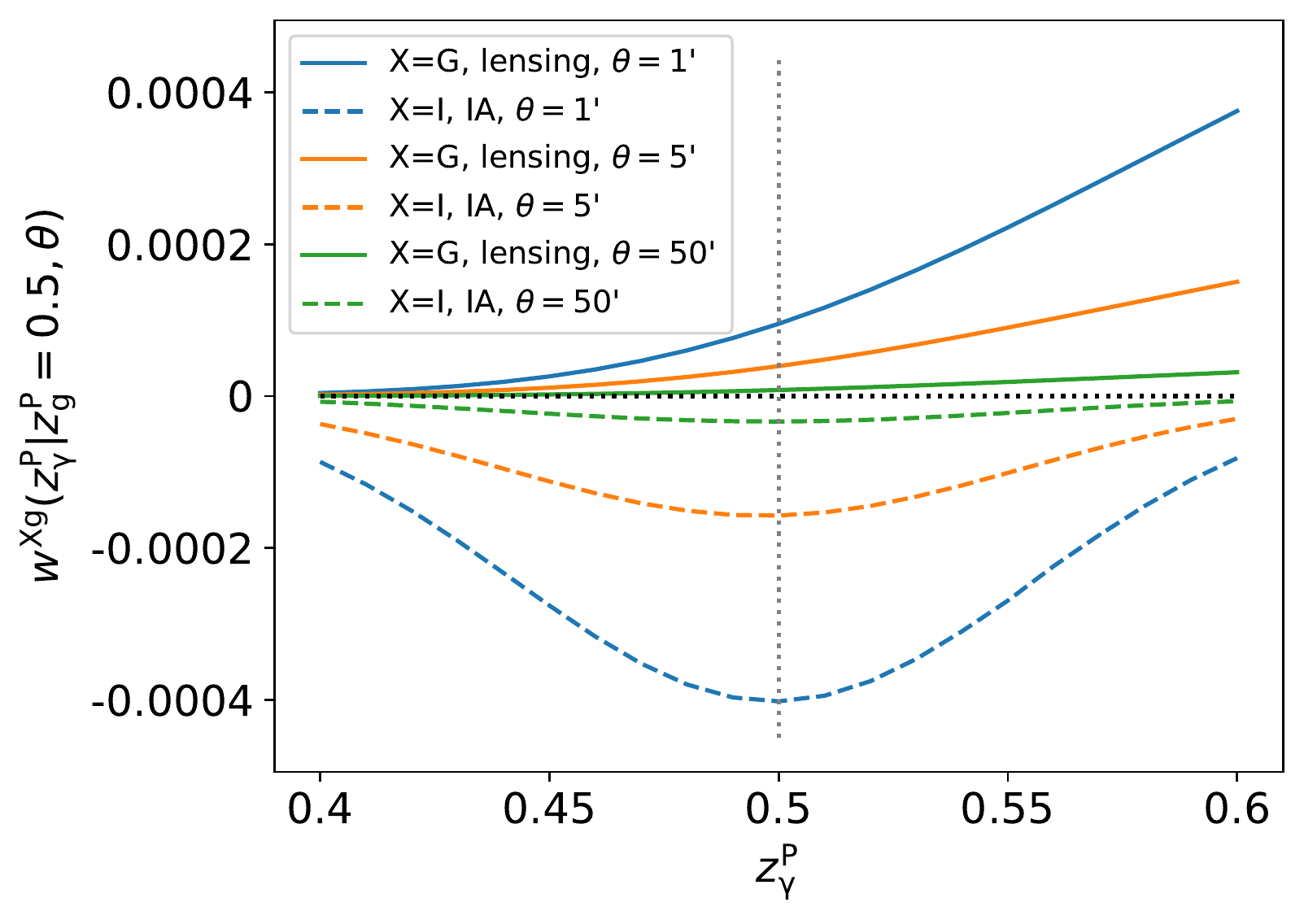}
        \caption{Toy model to illustrate the different redshift dependences for the lensing signal and the IA signal and to explain why the SC selection Eq.\,\ref{eq SC selection} works. We placed many lens galaxies at photo-z $z^{\rm P}_{\rm g}=0.5$ (the grey dotted line), while allowing the photo-z of the source galaxies $z^{\rm P}_{\rm \gamma}$ to change (x-axis) to evaluate the corresponding lensing correlation function $w^{\rm Gg}$ or IA correlation function $w^{\rm Ig}$ at different angular separation $\theta$. The true-z has a Gaussian scatter of 0.04 (this number is chosen for display purposes, so that the lensing/IA signals have comparable maximum and minimum values) around the photo-z, for both source galaxies and lens galaxies. As the gravitational lensing shear is an optical shape that requires $z_{\rm g}<z_{\rm \gamma}$, it will have a non-symmetric power around $z^{\rm P}_{\rm g}$, as the positive solid curves show. This also demonstrate $Q^{\rm Gg}\ll1$ according to Eq.\,\ref{eq Q^Gg theta}. As the IA shape is a dynamical shape, it does not have requirements on the relative redshifts, leading to a symmetric power around $z^{\rm P}_{\rm g}$, as the negative dashed curves show. This also demonstrate $Q^{\rm Ig}\sim1$ according to Eq.\,\ref{eq Q^Ig theta}. These relations hold for signals at different angular separations (different colors). The different IA models (which could deviate from Eq.\,\ref{eq IA P(k) model} and with $A_{\rm IA}=1$ being assumed) will only change the relative amplitudes of the negative signals at different scales, but not the redshift-dependency around $z^{\rm P}_{\rm g}$. We note that at such a redshift range, the magnification signal is much smaller than the IA signal.}
        \label{fig SC illustration}
\end{figure}

The two SC observables can be written as:
\begin{align}
        w^{\rm\gamma g^L}_{ii}(\theta)&=w^{\rm Gg}_{ii}(\theta)+w^{\rm Ig}_{ii}(\theta)+g_{\rm mag}\left[w^{\rm G\kappa^{\rm gal}}_{ii}(\theta)+w^{\rm I\kappa^{\rm gal}}_{ii}(\theta)\right], \label{eq w^gamma-g} \\
        w^{\rm \gamma g^L}_{ii}|_{\rm S} (\theta)&=w^{\rm Gg}_{ii}|_{\rm S} (\theta)+w^{\rm Ig}_{ii}|_{\rm S}(\theta) +g_{\rm mag} \left[w^{\rm G\kappa^{\rm gal}}_{ii}|_{\rm S}(\theta)+w^{\rm I\kappa^{\rm gal}}_{ii}|_{\rm S}(\theta) \right] \label{eq
                w^gamma-g|S}\ ,
\end{align}
where ``$|_{\rm S}$'' denotes the SC selection, and $i$ denotes the $i$-th redshift bin if tomography is applied. The lensing-galaxy $w^{\rm Gg}$ and the IA-galaxy $w^{\rm Ig}$ signal are affected by this SC selection, as quantified by the $Q$ parameters:
\begin{align}
        Q^{\rm Gg}_i(\theta) &\equiv \frac{w^{\rm Gg}_{ii}|_{\rm S}(\theta)}{w^{\rm Gg}_{ii}(\theta)}, \label{eq Q^Gg theta}\\
        Q^{\rm Ig}_i(\theta) &\equiv \frac{w^{\rm Ig}_{ii}|_{\rm S}(\theta)}{w^{\rm Ig}_{ii}(\theta)}. \label{eq Q^Ig theta}
\end{align}\\
For the lensing signal to exist, the redshift of the source, $z_\gamma$, needs to be greater than the redshift of the lens, $z_{\rm g}$: $z_\gamma>z_{\rm g}$. The SC photo-z selection $z^{\rm P}_\gamma<z^{\rm P}_{\rm g}$ largely reduces the lensing signal, leading to $Q^{\rm Gg}\ll1$. The IA signal does not rely on the ordering along the line of sight, with $Q^{\rm Ig}\sim1$. The lensing drop, $Q^{\rm Gg}$, and the IA drop, $Q^{\rm Ig}$, are dependent on the photo-z quality, as described in \cite{SC2008,Yao2017,Yao2019,Yao2020}. If the photo-z quality is perfect, the SC selection will result in no lensing signal so that $Q^{\rm Gg}$ approaches 0. For incorrect photo-zs, the SC selection fails and $Q^{\rm Gg}$ is $\sim1$. Given a photo-z distribution $n^{\rm P}(z^{\rm P})$ and the true-z distribution $n(z)$, the lensing-drop $Q^{\rm Gg}$ and IA-drop $Q^{\rm Ig}$ can be theoretically derived, following \cite{Yao2019,Yao2020}, with more technical details provided in Appendix \ref{Appendix Q}. We also present a toy model to visualize how the SC selection works in Fig.\,\ref{fig SC illustration}.

\begin{figure}\centering
        \includegraphics[width=1.0\columnwidth]{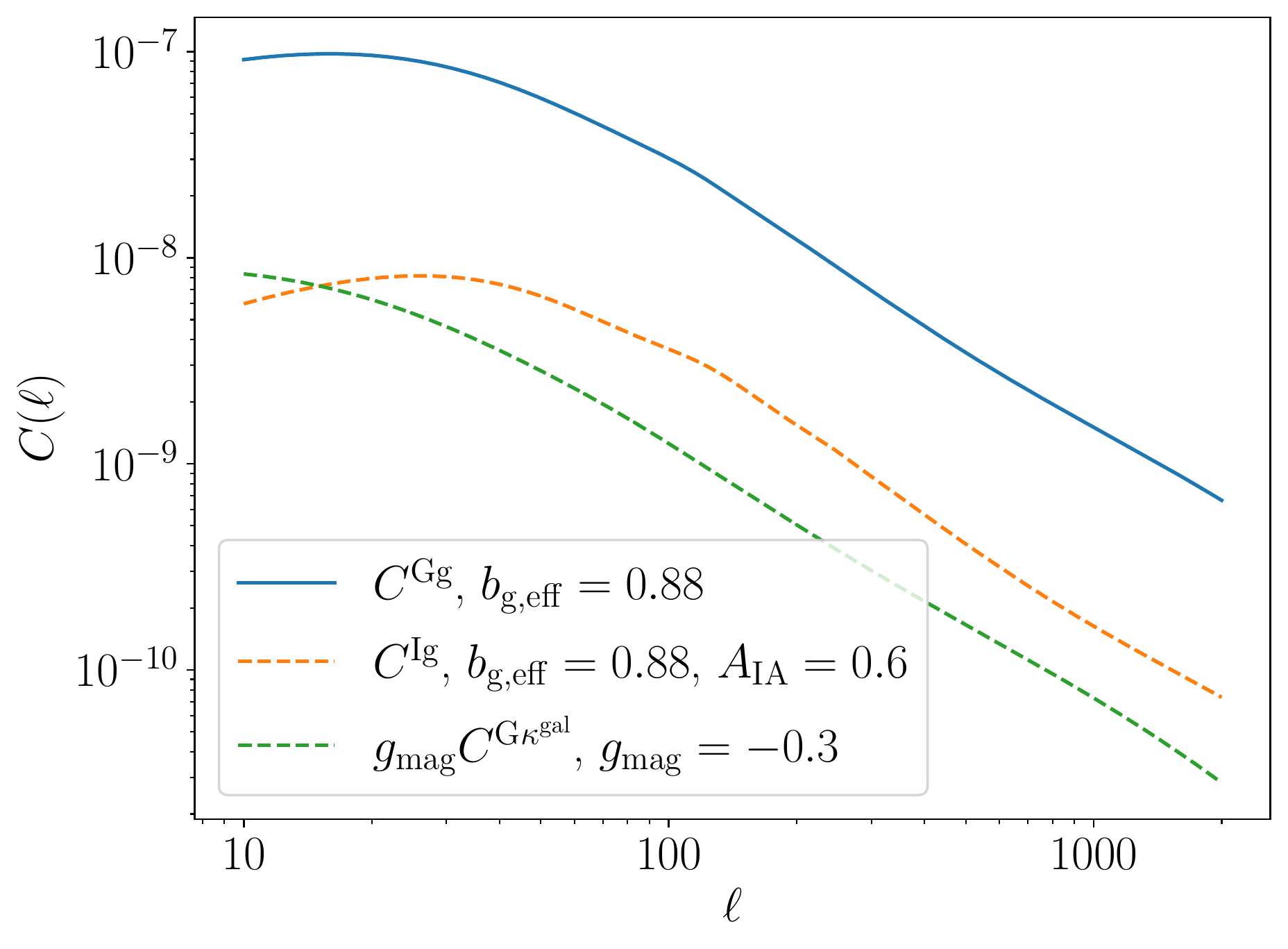}
        \caption{Theoretical comparison between the galaxy-shear $C^{\rm Gg}(\ell)$, galaxy-IA $C^{\rm Ig}(\ell)$ and shear-magnification $g_{\rm mag}C^{\rm G\kappa^{\rm gal}}(\ell)$ angular power spectra, with the best fit of our baseline analysis and the redshift distribution $n(z)$ from KiDS-1000 $0.5<z^{\rm P}<1.2$ shear catalog. The dashed lines represent negative signals. This figure demonstrates that the magnification contamination is important in the self-calibration method for the high-z KiDS source sample.}
        \label{fig Cell}
\end{figure}

We quantitatively test the terms in Eq.\,\ref{eq w^gamma-g}, finding they generally follow $|w^{\rm I\kappa^{\rm gal}}|<|w^{\rm G\kappa^{\rm gal}}| \ll |w^{\rm Ig}|<|w^{\rm Gg}|$ for $z<0.9$ data; therefore, in previous analyses \citep{SC2008,Yao2019,Yao2020}, the magnification terms were neglected. For the $z\sim1$ galaxies, however, the magnification term $w^{\rm G\kappa^{\rm gal}}$ quickly approaches $w^{\rm Ig}$ and becomes a non-negligible source of contamination to the SC method. In Fig.\,\ref{fig Cell}, we show a theoretical comparison of the angular power spectra. We can write the SC selection for the magnification term as $w^{\rm G\kappa^{\rm gal}}|_{\rm S} = Q^{\rm G\kappa}w^{\rm G\kappa^{\rm gal}}$. At the drop of the signal, $Q^{\rm G\kappa}\sim Q^{\rm Ig} \sim 1,$ and given that these are not $z$-pair-dependent correlations,  the magnification signal $w^{\rm G\kappa^{\rm gal}}$ will contaminate the IA signal $w^{\rm Ig}$ due to similar behavior, leaving the lensing signal $w^{\rm Gg}$ unaffected. We note the $w^{\rm I\kappa}$ term is negligible in this work.

After measuring the galaxy-galaxy lensing observables \{$w^{\rm \gamma g^L}$, $w^{\rm \gamma g^L}|_{\rm S}$\} and the drops of the signals, \{$Q^{\rm Gg}$, $Q^{\rm Ig}$\}, (more details are provided in Eqs.\,\ref{eq Q^Gg theta}
and \ref{eq Q^Ig theta} and Appendix\,\ref{Appendix Q}), the corresponding lensing-galaxy correlation $w^{\rm Gg}$, IA-galaxy correlation $w^{\rm Ig}$ and shear-magnification correlation $w^{\rm G\kappa}$ can be linearly obtained:
\begin{align}
        w^{\rm Gg}_{ii}(\theta)&=\frac{Q^{\rm Ig}_{i}(\theta)w^{\gamma \rm g^L}_{ii}(\theta) - w^{\rm \gamma g^L}_{ii}|_{\rm S}(\theta)}{Q^{\rm Ig}_i(\theta)-Q^{\rm Gg}_{i}(\theta)}, \label{eq Gg correlation}\\
        w^{\rm Ig}_{ii}(\theta)+w^{\rm G\kappa^{\rm gal}}_{ii}(\theta)&=\frac{w^{\rm \gamma g^L}_{ii}|_{\rm S}(\theta) -Q^{\rm Gg}_{i}(\theta) w^{\rm \gamma g^L}_{ii}(\theta)}{Q^{\rm Ig}_i(\theta)-Q^{\rm Gg}_{i}(\theta)} \label{eq Ig correlation}\ .
\end{align}

In the previous work, the IA information was directly extracted in $w^{\rm Ig}$. However, as shown in Fig.\,\ref{fig Cell} and Eq.\,\ref{eq Ig correlation}, for KiDS the subtracted signal suffers from the contamination from a magnification term, $w^{\rm G\kappa}$. Constraining the measurements of \{$w^{\rm Gg}$, $w^{\rm Ig}$+$w^{\rm G\kappa^{\rm gal}}$, $w^{\gamma\kappa^{\rm CMB}}$\} together, including the covariance, will lead to robust constraints on both the lensing amplitude and the nuisance parameters. For the current stage, where the S/N values for the measurements are not very high, we choose to ignore the possible scale-dependent features for the effective galaxy bias $b_{\rm g,eff}$ and IA amplitude $A_{\rm IA}$, and assume they are linear and deterministic. The parameters \{$A_{\rm lens}$, $A_{\rm IA}$, $b_{\rm g,eff}$, $g_{\rm mag}$\} are connected to the observables following:
\begin{align}
        w^{\rm Gg}(\theta) &= b_{\rm g,eff} w^{\rm Gm}_{\rm theory}(\theta), \label{eq w^Gg fit}\\
        w^{\rm Ig}(\theta)+w^{{\rm G}\kappa^{\rm gal}}(\theta) &= b_{\rm g,eff} A_{\rm IA} w^{\rm Im}_{\rm theory}(\theta) + g_{\rm mag}w^{{\rm G}\kappa^{\rm gal}}_{\rm theory}(\theta), \label{eq w^Ig fit}\\
        w^{\gamma\kappa^{\rm CMB}}(\theta) &= A_{\rm lens}w^{{\rm G}\kappa^{\rm CMB}}_{\rm theory}(\theta) + A_{\rm IA}w^{{\rm I}\kappa^{\rm CMB}}_{\rm theory}(\theta), \label{eq w^GK fit}\
\end{align}
where ``m'' stands for matter, which is the case if one sets the effective galaxy bias $b_{\rm g,eff}=1$. We separate the CMB convergence and the galaxy convergence (due to magnification) with $\kappa^{\rm CMB}$ and $\kappa^{\rm gal}$. On the LHS of Eq.\,\ref{eq w^Gg fit}, \ref{eq w^Ig fit}, and \ref{eq w^GK fit}, we have the measurements, while on the RHS the correlations $w(\theta),$  we have the theoretical predictions assuming {\it Planck} cosmology \citep{Planck2018I}, see Table \ref{table fiducial cosmology}. We note the $Q$ values being used to obtain the LHS are also cosmology-dependent, however, the sensitivity is weak as the cosmological part is mostly canceled when taking the ratio in Eq.\,\ref{eq Q^Gg theta} and \ref{eq Q^Ig theta}. We tested whether the fiducial cosmology is changed to any of the KiDS-1000 cosmologies in Table \ref{table fiducial cosmology}. We see the $Q$s will change by $\sim1\%$, similarly to what is found in \cite{Yao2020}, and the resulting changes to the fitting parameters \{$A_{\rm IA}$, $b_{\rm g,eff}$, $g_{\rm mag}$, $A_{\rm lens}$\} are negligible. However, considering the RHS, those four fitting parameters are sensitive to the fiducial cosmology used to produce the $w_{\rm theory}$ values when magnification exists, which differs from previous analysis \citep{Yao2020}. The theoretical predictions $w_{\rm theory}$ are calculated with \textsc{ccl}\footnote{Core Cosmology Library, \url{https://github.com/LSSTDESC/CCL}} \citep{Chisari2019CCL} and \textsc{camb}\footnote{Code for Anisotropies in the Microwave Background, \url{https://camb.info/}} \citep{Lewis2000CAMB}. The effective galaxy bias $b_{\rm g,eff}$ in this work is used to separate from the true galaxy bias of this sample, as we will discuss later it can absorb several sources of systematics.

\begin{table}
        \centering
        \caption{$\Lambda$CDM cosmological parameters adopted in this work, corresponding to the best-fit cosmology from \cite{Planck2018I}, and the KiDS-1000 multivariate maximum posterior (MAP) results from the two-point correlation functions $\xi_\pm$, the band powers $C(\ell)$, and the COSEBIs, as in \cite{Asgari2021}. \label{table fiducial cosmology}}    \begin{tabular}{ c c c c c c c c }
                \hline
                Survey & $h_0$ & $\Omega_{\rm b} h^2$ & $\Omega_{\rm c} h^2$ & $n_{\rm s}$ & $\sigma_8$ \\
                \hline
                {\it Planck} & 0.673  &  0.022  &  0.120  &  0.966  & 0.812 \\
                \hline
                KiDS $\xi_\pm$ & 0.711  &  0.023  &  0.088  &  0.928  & 0.895 \\
                \hline
                KiDS $C(\ell)$ & 0.704  &  0.022  &  0.132  &  0.999  & 0.723 \\
                \hline
                KiDS COSEBI & 0.727  &  0.023  &  0.105  &  0.949  & 0.772 \\
                \hline
        \end{tabular}
\end{table}

The theoretical prediction of $w^{\rm G\kappa^{\rm CMB}}_{\rm theory}(\theta)$ is given in Eq.\,\ref{eq w Hankel}, and $w^{\rm I\kappa^{\rm gal}}_{\rm theory}(\theta)$ is obtained similarly with the Hankel transform from its power spectrum as in Eq.\,\ref{eq C^IK}. The $w^{\rm Gm}_{\rm theory}$, $w^{\rm Im}_{\rm theory}$, and $w^{\rm G\kappa^{\rm gal}}_{\rm theory}$ terms are the Hankel transform from the following angular power spectra:
\begin{align}
        C^{\rm Gm}(\ell) &= \int_{z_{\rm min}}^{z_{\rm max}} \frac{q^{\rm gal}(\chi)n(\chi)}{\chi^2}P_\delta\left(k=\frac{\ell+1/2}{\chi},z\right)d\chi, \label{eq C^Gm} \\
        C^{\rm Im}(\ell) &= \int_{z_{\rm min}}^{z_{\rm max}} \frac{n(\chi)n(\chi)}{\chi^2}P_{\delta,\gamma^I}\left(k=\frac{\ell+1/2}{\chi},z\right)d\chi, \label{eq C^Im} \\
        C^{\rm G\kappa^{\rm gal}}(\ell) &= \int_{z_{\rm min}}^{z_{\rm max}} \frac{q^{\rm gal}(\chi)q^{\rm gal}(\chi)}{\chi^2}P_\delta\left(k=\frac{\ell+1/2}{\chi},z\right)d\chi.
\end{align}

As discussed in previous work \citep{Yao2020}, by including the effective galaxy bias, $b_{\rm g,eff}$, we can obtain an unbiased estimation of $A_{\rm IA}$. This information will be propagated into Eq.\,\ref{eq w^GK fit} to break the degeneracy between $A_{\rm IA}$ and $A_{\rm lens}$. In this work, we further extend the fitting to include the impact from magnification with the nuisance parameter, $g_{\rm mag}$. We show later in this work that an unbiased CMB lensing amplitude, $A_{\rm lens}$, can be obtained from the simultaneous fitting of Eqs.\,\ref{eq w^Gg fit}, \ref{eq w^Ig fit}, and \ref{eq w^GK fit}.

\section{Data} \label{Section data}

In this section, we introduce the data we use for the $\left<\gamma\kappa^{\rm CMB}\right>$ cross-correlation study. Additionally, we used mock KiDS data, based on the MICE2 simulation (see \cite{vandenBusch2020} for details) to quantify the potential bias in the SC method due to magnification, photo-z modeling, and the boost factor.

\subsection{KiDS-1000 shear catalog}

We used the fourth data release of the Kilo-Degree Survey that covers $1006\deg^2$, known as KiDS-1000 \citep{Kuijken2019}. It has images from four optical bands, $ugri,$ and five near-infrared bands, $ZYJHK_s$. The observed galaxies can reach a primary $r-$band median limiting $5\sigma$ point source magnitude at $\sim25$. The shear catalog \citep{Giblin2021} contains $\sim21$ M galaxies and is divided into five tomographic bins in the range $0.1<z_B<1.2$ based on the \textsc{bpz} \citep{Benitez2000} algorithm. The ellipticity dispersion $\sigma_\epsilon$ is $\sim0.27$ per component and the shear multiplicative bias is generally consistent with 0. The KiDS data were processed by \textsc{theli} \citep{Erben2013} and Astro-WISE \citep{Begeman2013,deJong2015}. Shears are measured using \textit{lens}fit \citep{Miller2013}, and photometric redshifts are obtained from PSF-matched photometry and calibrated using external overlapping spectroscopic surveys \citep{Hildebrandt2021}. 

\begin{figure}\centering
        \includegraphics[width=1.0\columnwidth]{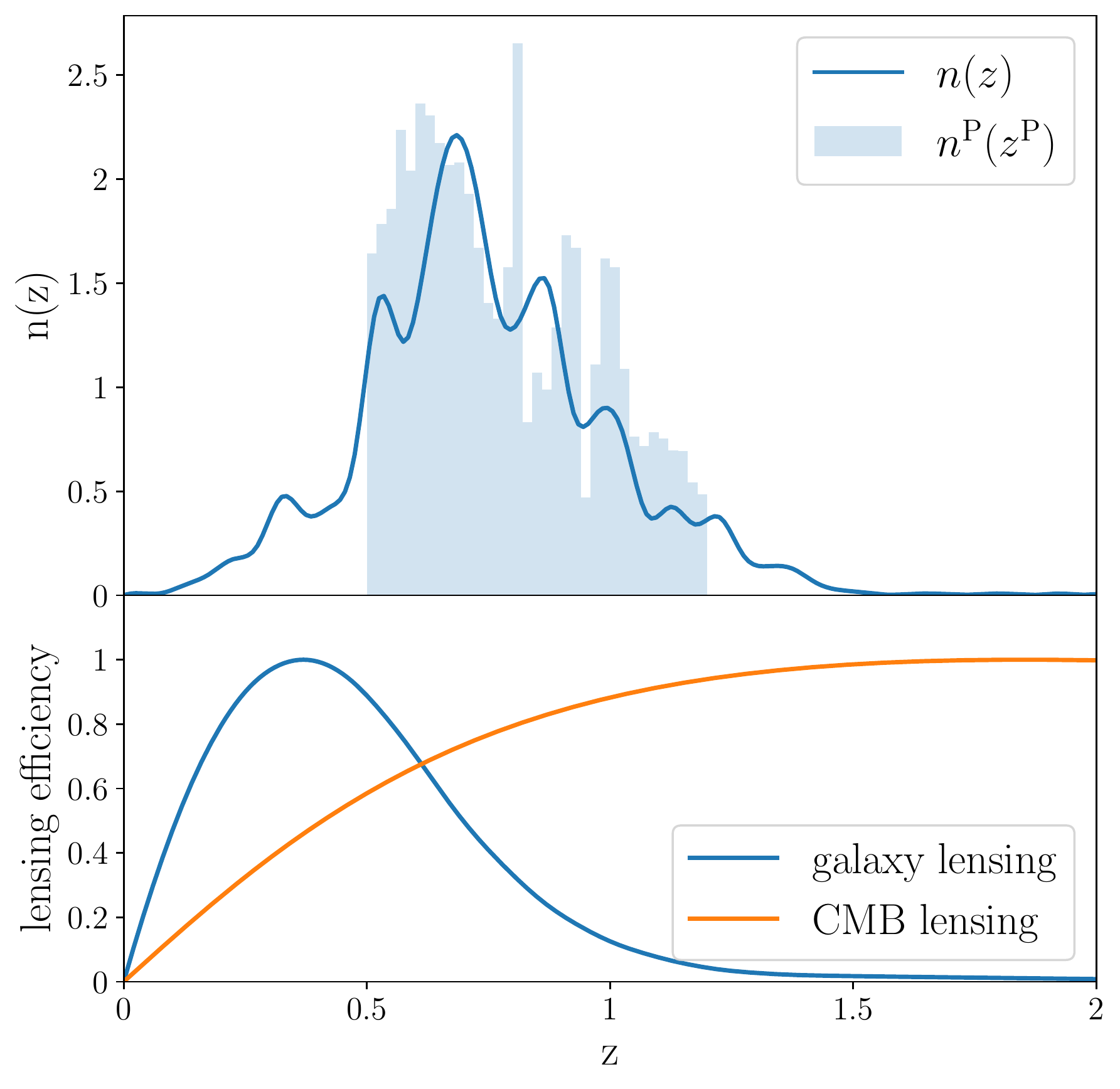}
        \caption{Photo-z distribution and the reconstructed redshift distribution of the combined galaxy sample in this work from self-organizing maps (SOM). The corresponding galaxy lensing efficiency Eq.\,\ref{eq q gal} and its comparison with CMB lensing efficiency Eq.\,\ref{eq q CMB} are shown in the lower panel, with the peak value normalized to 1.}
        \label{fig nz}
\end{figure}

The application of SC requires not only an accurate redshift distribution $n(z)$, but also relatively accurate photo-z for each galaxy, serving for the SC selection (Eq.\,\ref{eq SC selection}). We previously discussed \citep{Yao2019} the fact that the quality of photo-z is very important for the lensing-IA separation. Therefore in this work, we chose to combine the three high-z bins, namely, bin $3+4+5$ in the KiDS-1000 data, as a large bin so that the photo-z error for an individual galaxy is relatively small compared to the total bin width. The photo-z and the redshift distributions calibrated  with self-organizing maps (SOM) are shown in Fig.\,\ref{fig nz}. We chose to use the high-z bins because the CMB lensing efficiency Eq.\,\ref{eq q CMB} peaks at $z\sim1$ to $2$ (see lower panel of Fig.\,\ref{fig nz}), while the S/N for the cross-correlation is very low for the two low-z bins of KiDS-1000. 

To account for the selection functions for the shape of the footprint \citep{Mandelbaum2006} of the overlapped region and the varying galaxy number density due to observations \citep{Johnston2021J,Rezaie2020}, we divided the region into 200 sub-regions with a resolution of \textsc{Healpix} $N_{\rm side}=512$ ($\sim50$ arcmin$^2$ per pixel) and generated random points with 20 times the number of galaxies of the KiDS-1000 shear catalog in each sub-region. The pixels within the same sub-region are assigned the same galaxy numbers. This random catalog is used for the SC-related galaxy-galaxy lensing calculation, while its potential defects will not extend to cross-correlations.

\subsection{Planck legacy lensing map}

We use the CMB lensing map $\kappa(\vec{\theta})$ from the {\it Planck} data release \citep{Planck2018lensing}. The CMB lensing map is reconstructed with the quadratic estimator with the minimum-variance method combining the temperature map and the polarization map, after foreground removal with the SMICA method \citep{Planck2018I}. It covers $f_{\rm sky}=0.671$ of the whole sky with the maximum multiple $\ell=4096$.

In this work, we combine the footprint from the {\it Planck} lensing map and the mask of the KiDS-1000 shear catalog, leading to an overlapped region of $\sim829\deg^2$. We include the {\it Planck} Wiener filter \citep{Planck2018lensing}:
\begin{equation}
        \hat{\kappa}^{\rm WF}_{\ell m} = \frac{C^{\phi\phi, \rm fid}_\ell}{C^{\phi\phi, \rm fid}_\ell+N^{\phi\phi}_\ell} \hat{\kappa}^{\rm MV}_{\ell m} \label{eq Wiener filter}
,\end{equation}
to strengthen the CMB lensing signal at large scales, which will also lead to a suppression of the power spectrum at small scales, where the noise dominates \citep{Dong2021}. The Wiener filter 
is used both in the CMB lensing $\kappa$ map and in the theoretical predictions of Eq.\,\ref{eq C^GK} to prevent potential bias. After the application of the Wiener filter, we use \textsc{Healpy}\footnote{\url{https://github.com/healpy/healpy}} \citep{Healpy_Gorski2005,Healpy_Zonca2019} to convert the $\kappa_{\ell m}$ to the desired $\kappa$-map, and rotate from the galactic coordinates of {\it Planck} to the J2000 coordinates of KiDS with \textsc{Astropy} \citep{astropy}. The two-point correlation functions are calculated with \textsc{TreeCorr} \footnote{\url{https://github.com/rmjarvis/TreeCorr}} \citep{Jarvis2004}. 

\subsection{MICE2 mock catalog}

Additionally, we used the MICE2 simulation gold samples \citep{vandenBusch2020,Fosalba2015}, which highly mimic the KiDS-1000 shear catalog galaxies, to validate our SC method, concerning cosmic magnification and photo-z PDF model bias. MICE2 uses a simulation box width of $3.1~h^{-1}$Gpc, particle mass resolution of $2.9\times10^{10}~h^{-1}M_\odot$, and a total particle number of $\sim6.9\times10^{10}$. The fiducial cosmology is flat $\Lambda$CDM with $\Omega_{\rm m}=0.25$, $\sigma_8=0.8$, $\Omega_{\rm b}=0.044$, $\Omega_\Lambda=0.75,$ and $h=0.7$. The halos were identified with a friends-of-friends algorithm, as in \cite{Crocce2015}. The galaxies are populated within the halos with a mixture of halo abundance matching (HAM) and halo occupation distribution (HOD) up to $z\sim1.4$ \citep{Carretero2015}.

We note that in the MICE2 simulation that we use for the KiDS samples, intrinsic alignment is not yet included in the galaxy shapes (while an IA-included version can be found recently in \cite{Hoffmann2022}, but for DES\footnote{The Dark Energy Survey, https://www.darkenergysurvey.org/}, LOWZ \citep{Singh2016}, COSMOS\footnote{https://www.eso.org/qi/}, and other possible extensions). The fact that we aim to get $A_{\rm IA}=0$ to validate the SC method, while considering systematics from cosmic magnification and photo-z model bias, in addition to what has been addressed in \cite{Yao2020}. We used the galaxy positions (ra, dec), the two noiseless shear components ($\gamma_1$, $\gamma_2$), and BPZ-measured photo-z $z_B$ to calculate the SC correlations as in Eq.\,\ref{eq w^gamma-g} and \ref{eq w^gamma-g|S}. We tested the signal drop $Q$s of Eq.\,\ref{eq Q^Gg theta} and \ref{eq Q^Ig theta} with our photo-z PDF model and with true-z from simulation \citep{vandenBusch2020}. We compared the results using MICE2 gold samples (which highly mimic the KiDS-1000 shear catalog galaxies) with magnification (Eq.\,\ref{eq lensed overdensity}) and without magnification. For the MICE2 galaxies with magnification, we tested how it would bias the IA measurement and proved that when the magnification effect is also included in the model, IA can be measured in an unbiased way. The validations will be shown later in our results in more detail in Appendix \ref{Appendix Q}.

\section{Measurements} \label{Section measurements}

\begin{figure}\centering
        \includegraphics[width=1.0\columnwidth]{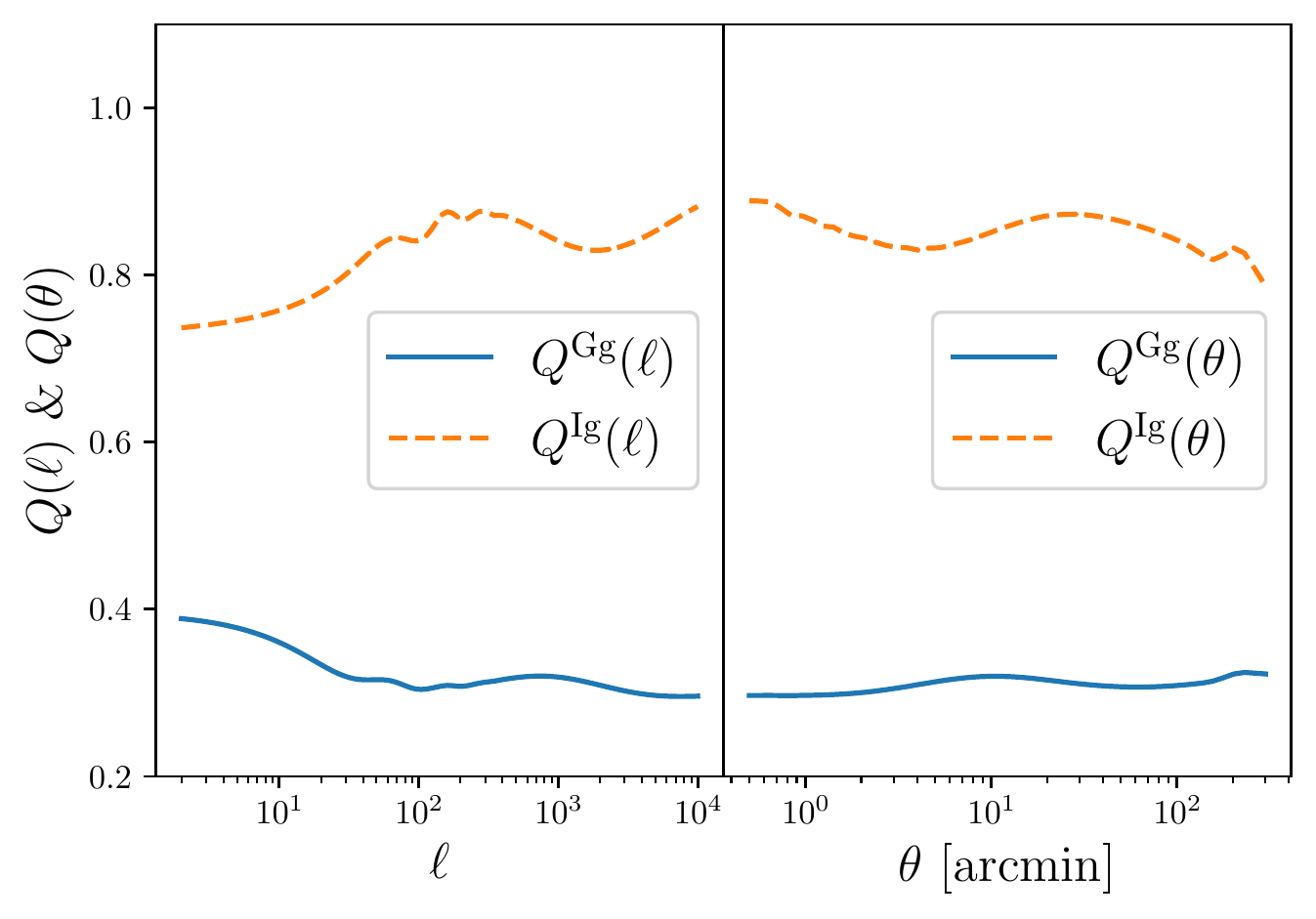}
        \caption{Lensing-drop $Q^{\rm Gg}$ and the IA-drop $Q^{\rm Ig}$ as a function of $\ell$ and $\theta$ by applying the SC selection Eq.\,\ref{eq SC selection};  see also Eqs.\,\ref{eq Q^Gg theta} and \ref{eq Q^Ig theta}. These values are adopted to obtain the separation of $w^{\rm Gg}$ and $w^{\rm Ig}+w^{\rm G\kappa^{\rm gal}}$, following Eqs.\,\ref{eq Gg correlation} and \ref{eq Ig correlation}. The left panel shows the calculation from power spectra and the right panel from correlation functions. The right panel is used to transfer \{$w^{\rm \gamma g}$, $w^{\rm \gamma g}|_S$\} to \{$w^{\rm Gg}$, $w^{\rm Ig}$\} later in Fig.\,\ref{fig GgIg}.}
        \label{fig Q}
\end{figure}

\begin{figure}\centering
        \includegraphics[width=1.0\columnwidth]{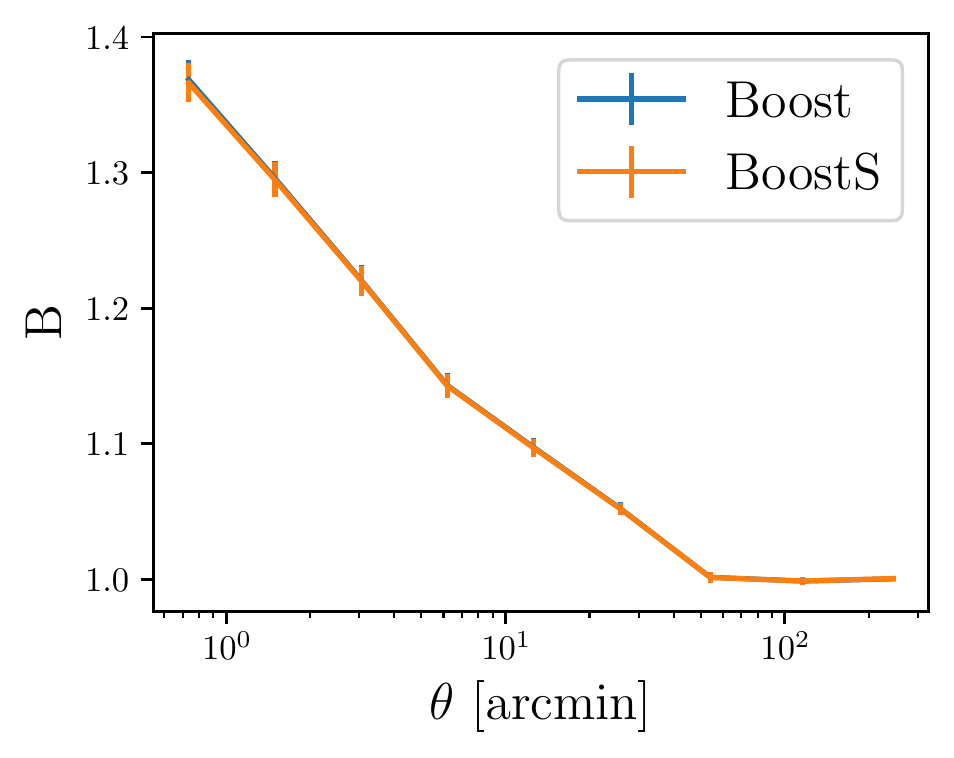}
        \caption{Boost factors for $w^{\rm\gamma g^L}$ and $w^{\rm\gamma g^L}|_{\rm S}$ are shown in blue and orange, respectively. Overlapping lines suggest the two signals are affected by the boost factor in almost the same way. We show the boost factor is significant at small scales for the SC observables. }
        \label{fig boost}
\end{figure}

\begin{figure*}\centering
        \includegraphics[width=2.0\columnwidth]{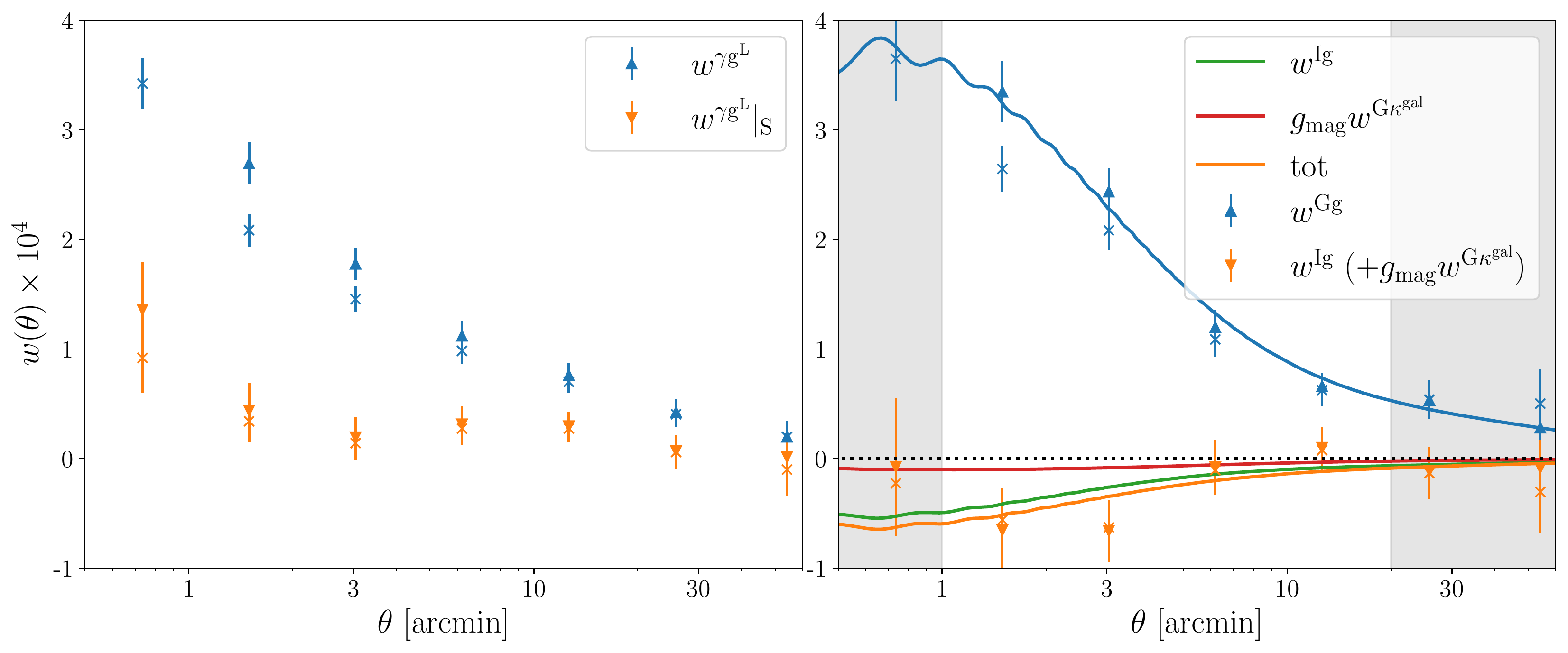}
        \caption{Measurements of SC. Left panel shows the measurement of the two introduced observables $w^{\rm\gamma g^L}$ and the one with the SC selection $w^{\rm\gamma g^L}|_{\rm S}$, while the corresponding $45$-deg rotation test is consistent with $0$ for both measurements. The significant separation of the two signals shows that SC is applicable. Right panel shows the separated lensing signal $w^{\rm Gg}$ and $w^{\rm Ig}$ (derived using Eqs.\,\ref{eq Gg correlation}, \ref{eq Ig correlation}, and Fig.\,\ref{fig Q}), where the latter is contaminated by the magnification signal as shown in Eq.\,\ref{eq Ig correlation}. 
                The up- and down-triangles are the results that take the boost factor (Fig.\,\ref{fig boost}) into consideration, while the crosses are the results that ignore this correction, setting $B=1$. The curves are the theoretical value with the best-fit \{$A_{\rm IA}$, $b_{\rm g,eff}$, $g_{\rm mag}$\} of this work. The blue curve represents the separated lensing signal as in Eq.\,\ref{eq w^Gg fit}. Orange curve represents the total contribution of IA and magnification as in Eq.\,\ref{eq w^Ig fit}.}
        \label{fig GgIg}
\end{figure*}

We show the estimation of the signal-drops for lensing and IA due to the SC selection (as in Eqs.\,\ref{eq Q^Gg theta} and \ref{eq Q^Ig theta}), namely, the lensing-drop $Q^{\rm Gg}$ and the IA-drop $Q^{\rm Ig}$ in Fig.\,\ref{fig Q}.
They are responsible for the lensing-IA separation later in Fig.\,\ref{fig GgIg}, following Eqs.\,\ref{eq Gg correlation} and \ref{eq Ig correlation}. 
We followed the processes in \cite{Yao2019,Yao2020} and adopted a bi-Gaussian photo-z probability distribution function (PDF) model with a secondary peak representing the photo-z outlier problem. We require the PDF model to have the same mean-$z$ as in Fig.\,\ref{fig nz}, while closest describing the projection from $n^{\rm P}(z^{\rm P})$ to $n(z)$.
We will also show, for the first time, how the assumed photo-z PDF model can affect the results in the next section, with more details shown in Appendix\,\ref{Appendix Q}.

We calculate the SC correlation function estimator,
\begin{equation} \label{eq gamma-g estimator}
        w^{\rm \gamma g}(\theta)=B(\theta)\frac{\sum_{\rm ED}\textsc{w}_j\gamma^+_j}{(1+\bar{m})\sum_{\rm
                        ED}\textsc{w}_j}-\frac{\sum_{\rm ER}\textsc{w}_j\gamma^+_j}{(1+\bar{m})\sum_{\rm
                        ER}\textsc{w}_j}\ ,
\end{equation}
to obtain the measurements of $w^{\rm \gamma g}$ and $w^{\rm \gamma g}|_{\rm S}$ from the tangential shear of each galaxy $\gamma_j^+$. Here we sum over the ellipticity-density pairs ($\sum_{\rm ED}$) and the ellipticity-random pairs ($\sum_{\rm ER}$) in an annulus centered on $\theta$, where the shear weight, $\textsc{w}_j$, of the $j$-th galaxy and the average multiplicative bias, $\bar{m,}$  are accounted for. The estimator is binned in angular $\theta$ space, with nine logarithmic bins from 0.5 to 300 arcmin. We used the averaged multiplicative bias $\bar{m}$ from averaging over the three z-bins, weighted by the effective galaxy number density. This gives $\bar{m}=-0.0036$.

We account for the impact of the boost factor \citep{Mandelbaum2005boostfactor,Singh2017,Joachimi2021}, which is $B$ in Eq.\,\ref{eq gamma-g estimator}. It is defined as:
\begin{equation}
        B(\theta)=\frac{\sum_{\rm ED} \textsc{w}_j}{\sum_{\rm RD} \textsc{w}_j}, \label{eq boost}
\end{equation}
which is used to quantify the small-scale bias due to the clustering of lens galaxies and source galaxies \citep{Bernardeau1998,Hamana2002,Yu2015}. This is because when applying the weight-correction to minimize the impact from shape noise, the summation over the pairs, $\sum_{\rm ED}$, in the denominator of Eq.\,\ref{eq gamma-g estimator} can contain galaxy clustering information when the true-z of the source and lens overlaps. If $\textsc{w}_j=1$, the summation of $\sum_{\rm ED}$ is propotional to $1+w^{\rm gg}$, where $w^{\rm gg}$ is the source-lens clustering correlation function. Such a bias will lead to a scale-dependent systematics, as the galaxy clustering correlation function is larger at small scales. We show the measurements of the boost factor for $w^{\rm \gamma g^L}$ and $w^{\rm \gamma g^L}|_{\rm S}$ as in Eq.\,\ref{eq w^gamma-g} and \ref{eq w^gamma-g|S} and in Fig.\,\ref{fig boost}. The fact that the boost factors for $w^{\rm \gamma g^L}$ and $w^{\rm \gamma g^L}|_{\rm S}$ are identical suggests this bias can be absorbed by the galaxy bias $b_{\rm g,eff}$ parameter if magnification is absent ($g_{\rm mag}=0$), leading to an unbiased $A_{\rm IA}$ and $A_{\rm lens}$. The impact from the boost factor can potentially break the linear galaxy bias assumption, but later in Fig. \ref{fig GgIg}, we show the linear assumption is fine. The impact of the boost factor and magnification existing together is shown later in this paper.

In Fig.\,\ref{fig GgIg}, we show the SC measurements. In the left panel, the measured shape-galaxy correlations $w^{\rm \gamma g^L}$ are shown in blue: (1) the boost factor-ignored case ($B=1$) is shown as blue crosses, while (2) the boost factor-corrected case is shown as blue up-triangles. With the SC selection Eq.\,\ref{eq SC selection}, requiring $z^{\rm P}_\gamma<z^{\rm P}_{\rm g}$ for each galaxy pair, the lensing component will drop to $Q^{\rm Gg}\sim0.3$ and the IA component will drop to $Q^{\rm Ig}\sim0.85$ (for more details on $Q^{\rm Gg}$ and $Q^{\rm Ig}$, see Fig.\,\ref{fig Q} and Appendix\,\ref{Appendix Q}). Therefore, the selected correlations $w^{\rm \gamma g }|_{\rm S}$ will drop to the orange down-triangles. Similarly, the boost factor-ignored case is shown as crosses.

The separated lensing-galaxy signal $w^{\rm Gg}$ and IA-galaxy signal $w^{\rm Ig}$ (which is contaminated by magnification-shear signal $g_{\rm mag}w^{\rm G\kappa}$) are shown in the right panel of Fig.\,\ref{fig GgIg}. The blue and orange curves are the theoretical predictions with the best-fit \{$A_{\rm IA}$, $b_{\rm g,eff}$, $g_{\rm mag}$\}. For the fitting, we cut off the shaded regions at both large scales and small scales. The small scale cut at $\theta=1$ arcmin is based on the linear galaxy bias assumption, as including the $\theta<1$ arcmin data will make the fitting significantly worse (increasing the fitting $\chi^2$ from 7.5 to 50, with degree-of-freedom changed from 8 to 10). We note this scale cut could include the impacts from the 3D non-linear galaxy bias \citep{Fong2021} and other small-scale effects such as massive neutrinos or baryon feedback in the matter power spectrum \citep{Hildebrandt2016,Asgari2021}. We emphasize that these systematics will be absorbed by the effective galaxy bias parameter $b_{\rm g,eff}$ (without breaking the scale-independent bias assumption) so that the IA amplitude will not be affected. As discussed previously in \cite{Yao2019,Yao2020}, the SC method requires significant separation between $w^{\rm \gamma g^L}$ and $w^{\rm \gamma g^L}|_{\rm S}$ to accurately get $w^{\rm Gg}$ and $w^{\rm Ig}$. Therefore, we introduced a large-scale cut at $\theta=20$ arcmin due to insufficient separation for the left panel of Fig.\,\ref{fig GgIg}.

\begin{figure}\centering
        \includegraphics[width=1.0\columnwidth]{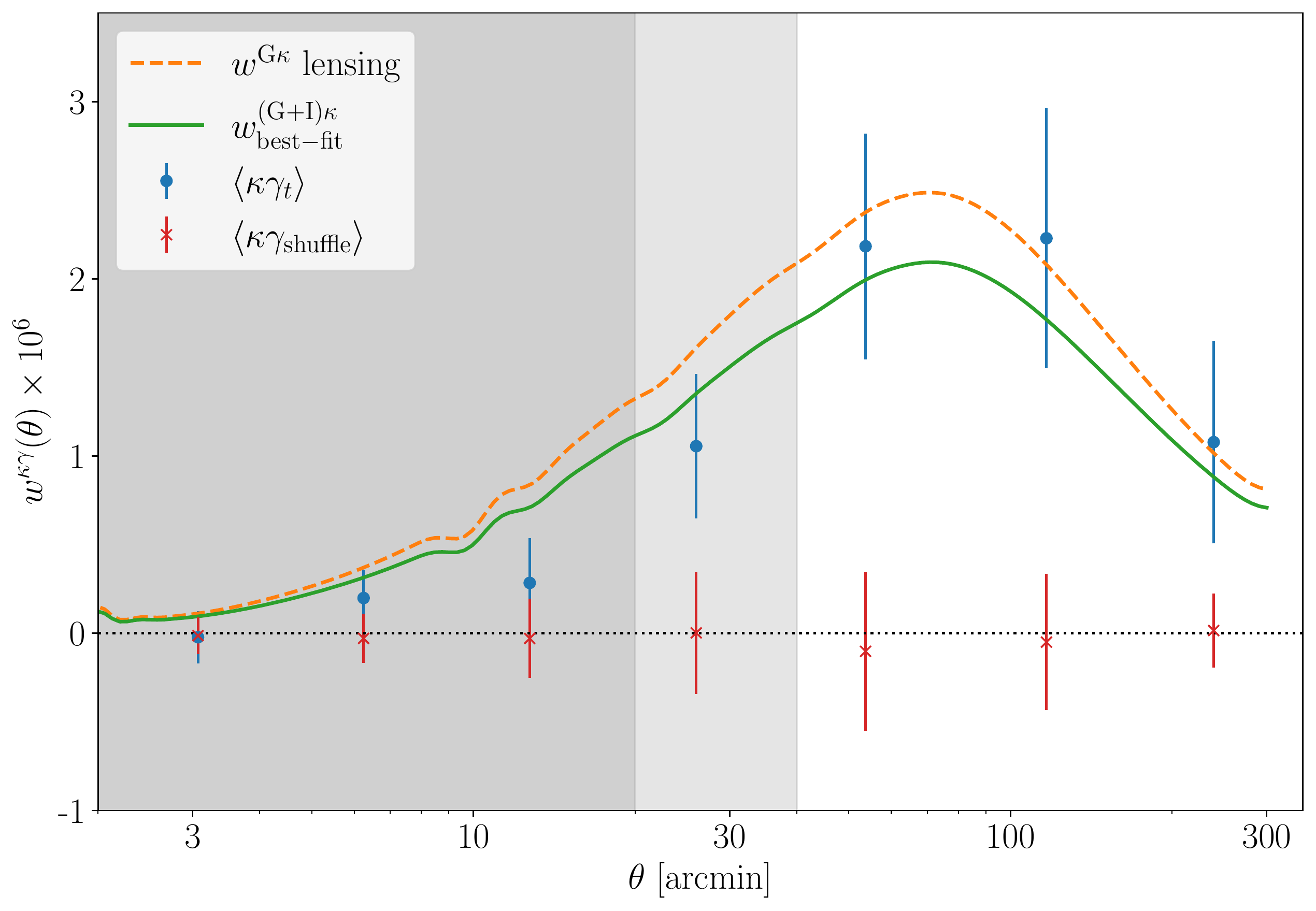}
        \caption{Measurement of the cross-correlation between {\it Planck} convergence $\kappa$ and KiDS-1000 shear $\gamma$, based on Eq.\,\ref{eq w^GK fit}. The blue dots are the measurements using tangential shear, with the green curve showing the best-fit considering both lensing and IA, while the orange curve shows only the lensing-lensing component. The red crosses show the null test by randomly shuffling the shear galaxies. The $45$-deg rotation tests for both the blue dots and the red dots are consistent with 0. The differently shaded regions correspond to our angular scale cuts at 2, 20 (default), and 40 arcmin.}
        \label{fig GK}
\end{figure}

Similarly, we measure the $\left<\gamma\kappa\right>$ correlation with the estimator:
\begin{equation}
        w^{\gamma\kappa}(\theta)=\frac{\sum_{ij} \textsc{w}_j\gamma^+_j\kappa_i}{(1+\bar{m})\sum_{ij} \textsc{w}_j},
\end{equation}
where $\kappa_i$ is the CMB lensing convergence in the $i$-th pixel of the pixelized map, taking the pixel center for its (ra, dec) coordinates, with $n_{\rm side}=2048$ in Healpy. The measured $w^{\gamma\kappa}$ values are shown in Fig.\,\ref{fig GK}. The tangential shear is shown as blue dots. We also show the measurements with randomly shuffling galaxy positions and the shear in red crosses as a null test. We tested the $45\deg$ rotated cross-shear for both the above cases and they are consistent with zero. The theoretical prediction with the best-fit $A_{\rm lens}$ and $A_{\rm IA}$ are shown as the green curve. If one assumes there is no IA in the measurements and uses $A_{\rm IA}=0$, the theoretical values for the pure lensing signal are shown in orange.

We note in Fig.\,\ref{fig GK} that because we use the Wiener-filtered $\kappa$ map from {\it Planck}, both the $w^{\gamma\kappa}$ measurements and the theoretical predictions are suppressed at small scales. The Wiener filter can significantly reduce the impact of the noise of the {\it Planck} lensing map and improve the S/N of the measurements.

\begin{figure}\centering
        \includegraphics[width=1.0\columnwidth]{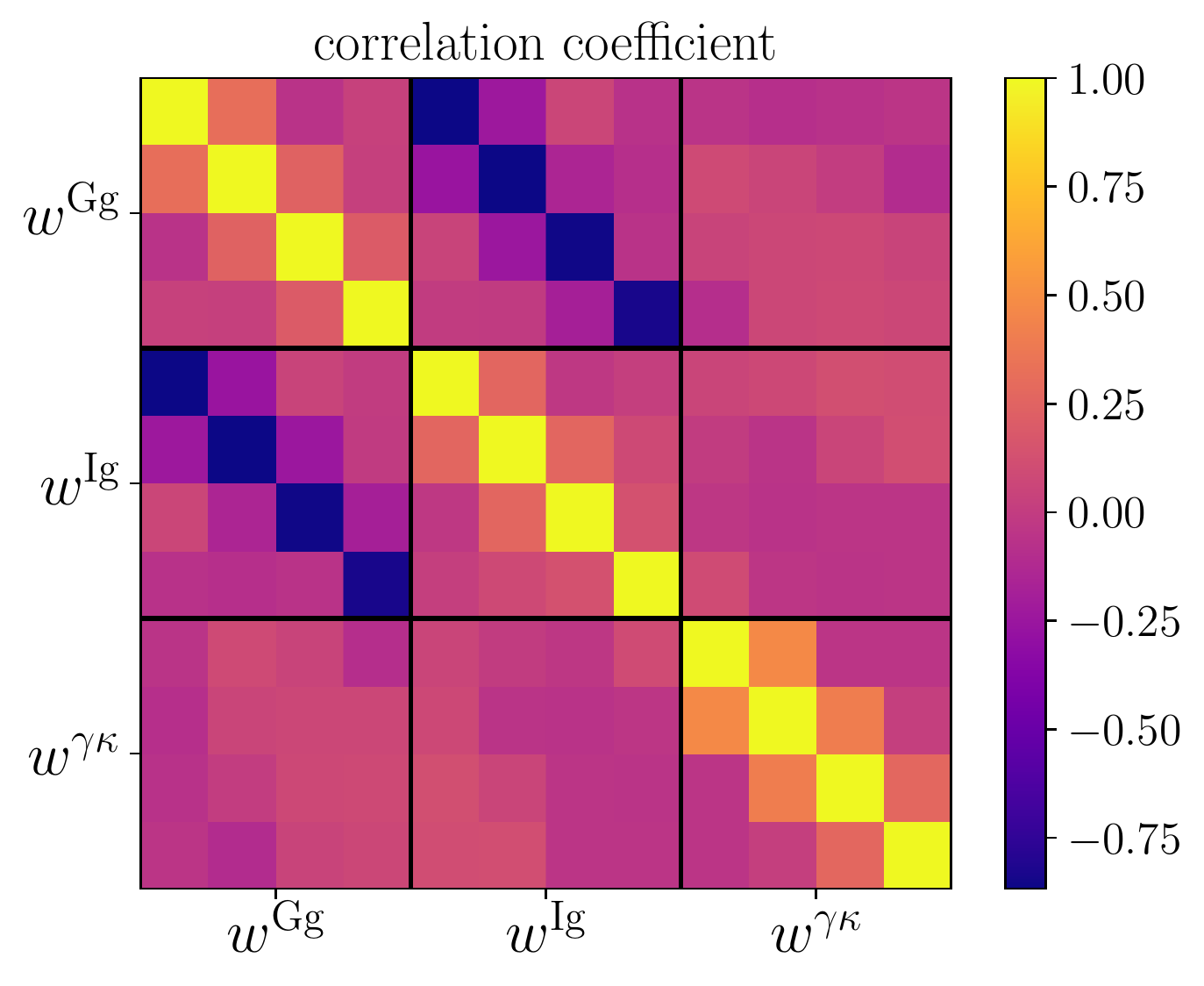}
        \caption{Normalized covariance matrix (i.e., the correlation coefficient) used in this work. There is a strong anti-correlation that exists between the lensing-galaxy correlation $w^{\rm Gg}$ and the IA-galaxy correlation $w^{\rm Ig}$ (including the contamination from $w^{\rm G\kappa^{\rm gal}}$), as we found in our previous work. The covariance of the 12 data points is calculated from Jackknife resampling with 200 regions. We note the IA information is passed from $1<\theta<20$ [arcmin] for $w^{\rm Ig}$ to $20<\theta<300$ [arcmin] for $w^{\rm \gamma\kappa}$ with the scale-independent $A_{\rm IA}$ assumption.}
        \label{fig cov}
\end{figure}

Together with the measurements in Figs.\,\ref{fig GgIg} and \ref{fig GK}, we obtain the observables of this work, which are the LHS terms of Eqs.\,\ref{eq w^Gg fit}, \ref{eq w^Ig fit}, and \ref{eq w^GK fit}. We used Jackknife resampling to obtain the covariance. $200$ Jackknife regions are used, which is much larger than the length of the data vector ($12$), based on the analysis of \cite{Mandelbaum2006,Hartlap2007}. The Jackknife regions are separated using the K-means algorithm \textsc{kmeans\_radec}\footnote{\url{https://github.com/esheldon/kmeans_radec}}. The normalized covariance matrix is shown in Fig.\,\ref{fig cov}. We find strong anti-correlation between $w^{\rm Gg}$ and $w^{\rm Ig}$ as expected \citep{Yao2020}. We note here in Fig.\,\ref{fig cov}, $w^{\rm Ig}$ means the separated signal in the RHS of Eq.\,\ref{eq Ig correlation}, including both the IA part and the contamination from magnification. There is no significant correlation between $w^{\gamma\kappa}$ and the other two observables. This covariance will be used in the Markov chain Monte Carlo  (MCMC) to find the best-fit parameters of \{$A_{\rm IA}$, $b_{\rm g,eff}$, $g_{\rm mag}$, $A_{\rm lens}$\}, while all the other cosmological parameters are fixed to Planck as in Table \ref{table fiducial cosmology}.

\section{Results} \label{Section results}

\subsection{Validation with MICE2}

In this subsection, we apply the IA self-calibration to the MICE2 mock catalog to test the impact of the systematics and validate the recovery of the IA signal. The processes of the mock data are identical to the descriptions in Sect.\,\ref{Section measurements}, but only focusing on the self-calibration part. The measurements are similar to those of Fig.\,\ref{fig GgIg}, so we chose to skip them. We performed the MCMC calculation using emcee \citep{emcee}. We considered flat priors in $-5<A_{\rm IA}<5$, $0<b_{\rm g,eff}<2,$ and $-3<g_{\rm mag}<3$.

\subsubsection{Impact from magnification}

\begin{figure}\centering
        \includegraphics[width=1.0\columnwidth]{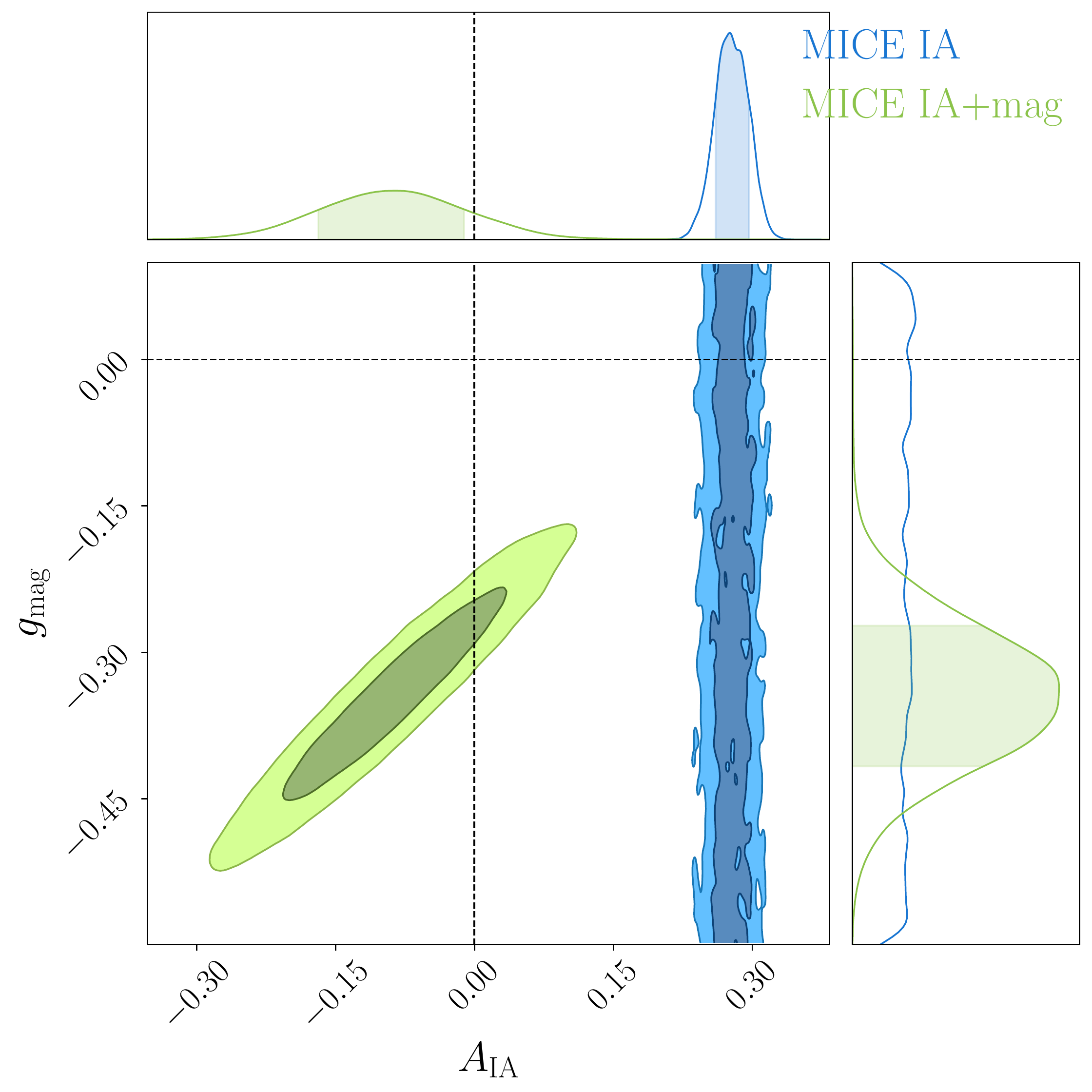}
        \caption{Impact of the magnification signal on the IA measurement in MICE2. The green and blue contours are with and without magnification models, respectively. 
                If the magnification model is used in the fitting, as in green, the IA amplitude $A_{\rm IA}$ is consistent with 0, which is the MICE2 input.}
        \label{fig MICE mag}
\end{figure}

We show how the magnification signal affects the original SC method \citep{SC2008,Yao2019,Yao2020} and the correction introduced in this work, focusing on the $g_{\rm mag}-A_{\rm IA}$ space.

In Fig.\,\ref{fig MICE mag}, we show that if magnification is not included in the modeling, $g_{\rm mag}$ is therefore not constrained. The existing magnification signal will be treated as the IA signal, leading to a non-vanishing $A_{\rm IA}\sim0.3$, which significantly deviates from the MICE2 input $A_{\rm IA}=0$. When the magnification model is included in the analysis, $A_{\rm IA}$ is then consistent with 0. This demonstrates the importance of including the magnification model in the SC analysis with high-z data. The results are also summarized later in the comparisons in Fig.\,\ref{fig compareAIA_MICE} for MICE2, and in Fig.\,\ref{fig compareAIA} for KiDS data.

We note that in the green case of Fig.\,\ref{fig MICE mag} that considered both IA and magnification, $g_{\rm mag}$ and $A_{\rm IA}$ strongly degenerate. Therefore the constraining power in $A_{\rm IA}$ has a significant loss compared with the blue case, which ignores magnification. This degeneracy can be broken in the future with higher S/N in the observables. This is because the shape of $w^{\rm Ig}$ and $w^{\rm G\kappa}$ are different at small scales for correlation functions as in Fig.\,\ref{fig GgIg}, and on large scales for power spectra as in Fig.\,\ref{fig Cell}. The IA-model dependency will be discussed later with other results. Based on the above analysis, we draw the conclusion that it is important to include magnification modeling for SC when using high-z data.

\subsubsection{Impact from modeling $p(z|z^{\rm P})$}

Since the SC selection Eq.\,\ref{eq SC selection} plays an important role in the lensing-IA separation process, it is crucial to understand how the following aspects affect SC: (1) the quality of the photo-z $z^{\rm P}$, (2) the true redshift distribution $n(z)$, and (3) the link between them $p(z|z^{\rm P})$. The quality of photo-z and the reconstruction of $n(z)$ has been studied thoroughly for KiDS data \citep{Kuijken2019,vandenBusch2022,Hildebrandt2021,vandenBusch2020}; therefore, we trust these results and leave the alternative studies for SC to future works. The uncalibrated PDF that projects $z^{\rm P}\rightarrow z$, on the other hand, has some known problems, for example, when Probability Integral Transform (PIT) is applied \citep{Newman2022,Hasan2022}.

In this work, we use a bi-Gaussian PDF model to project the photo-z distribution $n^{\rm P}(z^{\rm P})$ to the SOM redshift distribution $n(z)$, which are  shown in Fig.\,\ref{fig nz}. This modeling ignores the potential differences for galaxies in the same $z$-bin \citep{Peng2022,Xu2023}. However, this is an alternative process, considering the PDF problem for a single galaxy. This analytical approach is also much faster in calculation than using different PDFs for different galaxies.

\begin{figure}\centering
        \includegraphics[width=1.0\columnwidth]{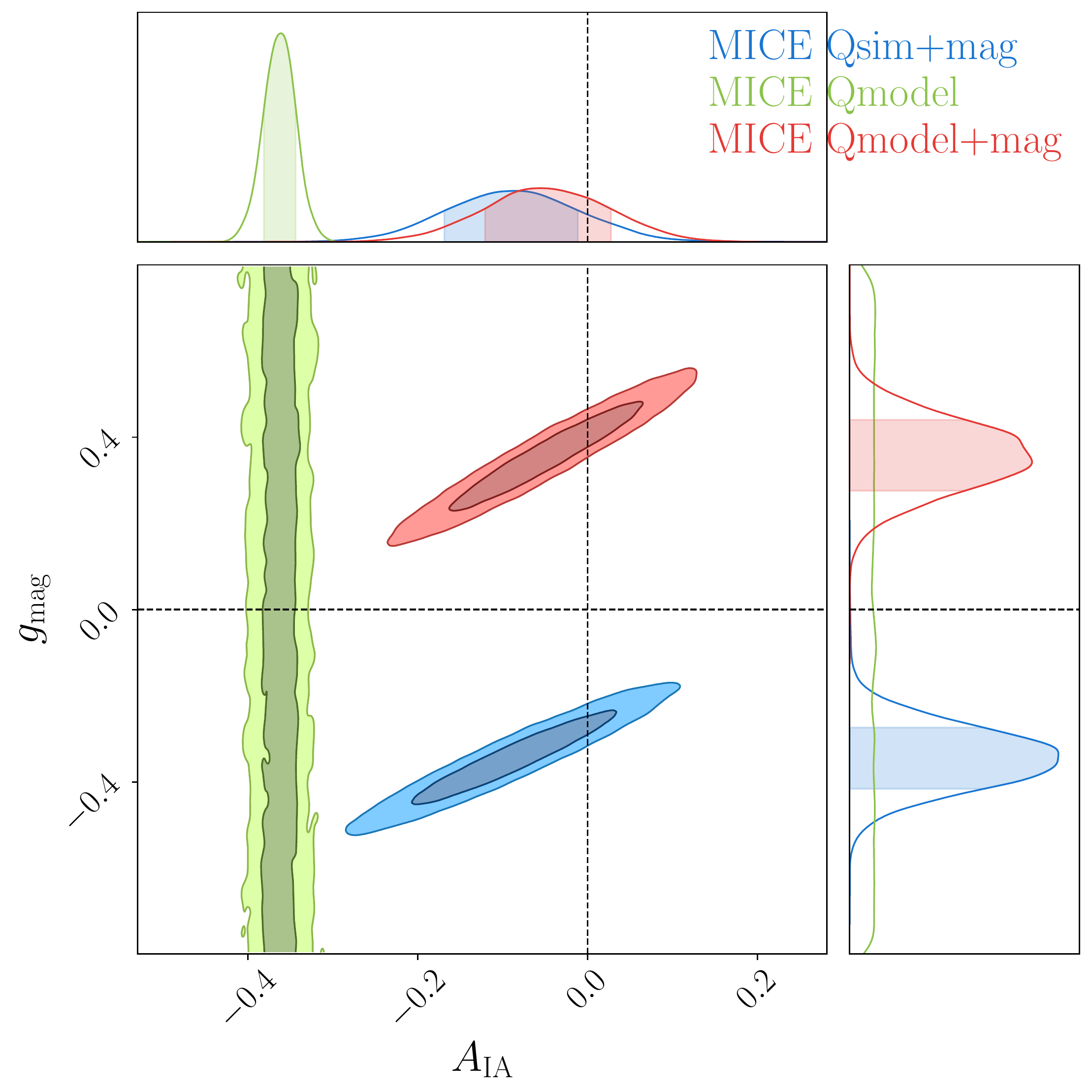}
        \caption{Impact from photo-z PDF model bias. The blue case uses photo-z from the BPZ algorithm and true-z for each galaxy to calculate Eq.\,\ref{eq K kernel} and the resulting $Q^{\rm Gg}$ and $Q^{\rm Ig}$, which are the ``sim'' cases in Fig.\,\ref{fig Q MICE}. This  $A_{\rm IA}$ is consistent with 0, which is the MICE2 input. The green case uses the bi-Gaussian photo-z model for the calculation, which are the ``model'' cases in Fig.\,\ref{fig Q MICE}, while ignoring the magnification contribution. This lead to unconstrained $g_{\rm mag}$ and biased $A_{\rm IA}$. In the red case, which also uses the photo-z model, but includes the magnification model, the resulting $A_{\rm IA}$ is still consistent with 0, with the bias from photo-z model error absorbed by $g_{\rm mag}$.}
        \label{fig MICE Qmodel}
\end{figure}

\begin{figure}\centering
        \includegraphics[width=1.0\columnwidth]{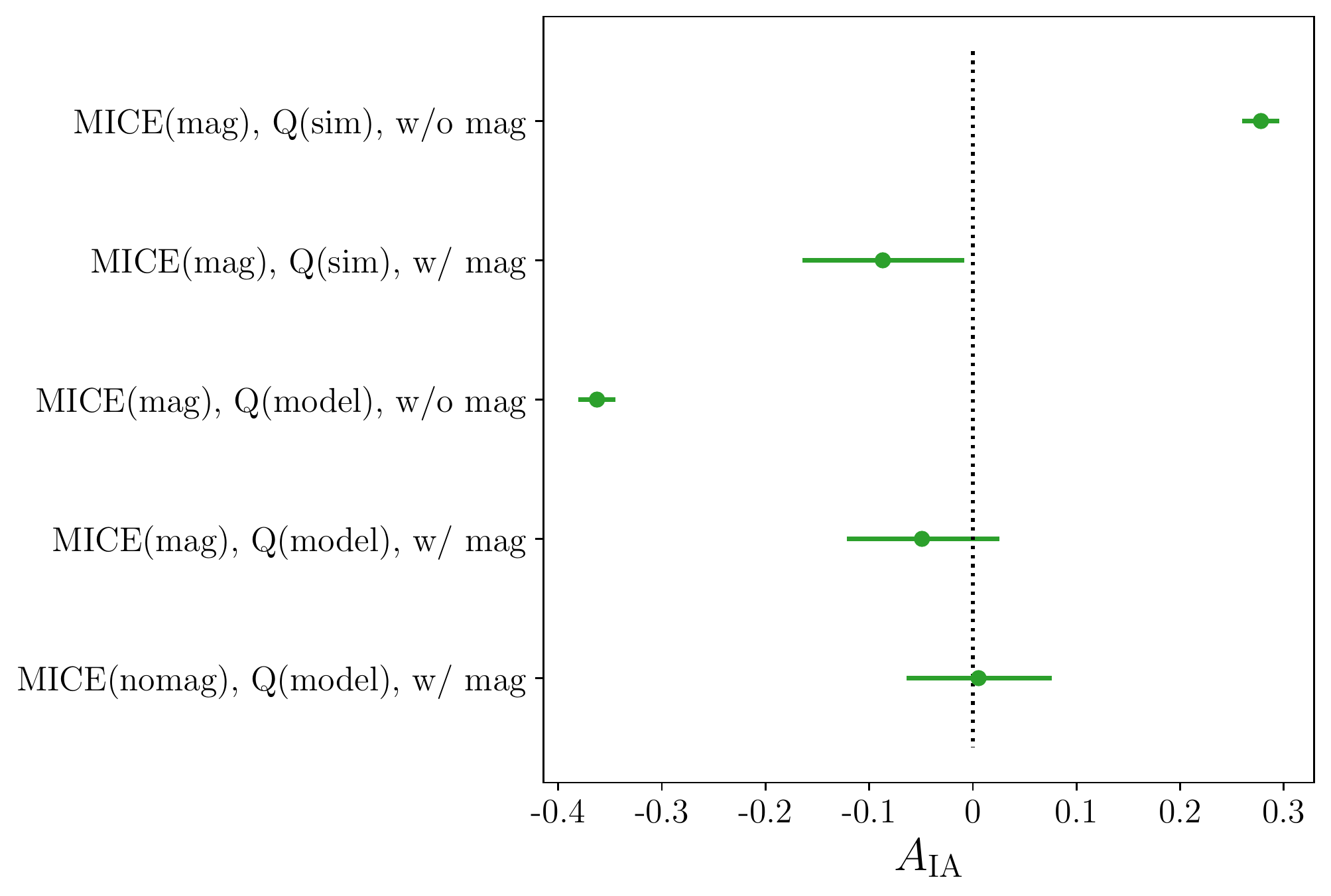}
        \caption{We validate our SC method with MICE2 simulation, which does not have IA implemented; therefore, $A_{\rm IA}=0$ is expected. The results are shown in green, with ``MICE(mag)'' meaning magnification is included in the MICE simulation, while ``MICE(nomag)'' means magnification is not included, ``Q(sim)'' and ``Q(model)'' mean if the signal drops $Q$ values are calculated from true-z from simulation or photo-z PDF model, and ``w/o mag'' and ``w/ mag'' show if the case includes magnification model in the fitting process. The upper two data points are the results from Fig.\,\ref{fig MICE mag}, showing the impact of the modeling magnification. The second to fourth data points are the results from Fig.\,\ref{fig MICE Qmodel}, showing the impact of Q calculation using different PDFs. The fourth data point corresponds to our fiducial analysis carried out later for the KiDS data, with potential bias $\Delta A_{\rm IA}<0.1$. The bottom data is a reference case assuming no magnification effects in the data, corresponding to our previous work \cite{Yao2020,Yao2019}.}
        \label{fig compareAIA_MICE}
\end{figure}

We used Fig.\,\ref{fig MICE Qmodel} to demonstrate how large this photo-z PDF modeling bias is with different approaches. We used a MICE2 simulation with galaxy number density affected by magnification. When the SC calculation uses true-z to calculate the signal drops $Q^{\rm Gg}$ and $Q^{\rm Ig}$, and the magnification model is also considered, we find the resulting $A_{\rm IA}$ is consistent with 0, which is the MICE2 input. The scatter on $A_{\rm IA}$ is $\sim0.1$, thanks to the noiseless shapes in MICE2. If the $Q$s are calculated with the assumed photo-z PDF model (based on Fig.\,\ref{fig nz}), without including the magnification model, then $A_{\rm IA}$ will be biased towards the negative direction. We proved with our fiducial analysis that, even if there a bias in $Q^{\rm Gg}$ exists due to the assumed photo-z model, as long as the magnification model is used, this bias will be absorbed by the $g_{\rm mag}$ parameter, so that the IA amplitude $A_{\rm IA}$ is unbiased (consistent with 0 in the MICE2 case). The results are also shown later in the comparisons in Fig.\,\ref{fig compareAIA_MICE} for MICE2, and in Fig.\,\ref{fig compareAIA} for KiDS data.

We note that it is not guaranteed that the bias coming from photo-z modeling will be absorbed by the magnification parameterization for all data sets. The above tests only validate this approach with KiDS or KiDS-like data that the residual bias $|\Delta A_{\rm IA}|<0.1$. However, it is not an essential problem for SC. In the future, if the photo-z outlier problem (or the redshift-color degeneracy problem) can be understood better, then a more reliable photo-z model can be used for our SC study. Alternatively, if the photo-z algorithms can give unbiased PDFs for each galaxy, this problem can also be directly solved. Direct approaches such as multiple SOM $n(z)$ for high-resolution photo-z bins (rather than the total five bins in the current KiDS data \cite{Asgari2021}) or using galaxy clustering to derive a scatter matrix to describe the PDF $p(z|z^{\rm P})$ \citep{Xu2023} can be investigated in the future. The combination between a color-based method and a clustering-based method \citep{Alarcon2020} can be even more promising in improving this aspect.

\subsection{Inference on real data}

With the above demonstration that our treatments for magnification and photo-z PDF are appropriate and the resulting bias in $A_{\rm IA}$ is very small, with ($|\Delta A_{\rm IA}|<0.1$ and $<1\sigma$ as shown in Fig.\,\ref{fig compareAIA_MICE}), we go on to apply SC to KiDS data and its cross-correlation with Planck lensing. We show the analysis of the following three situations:
(1) case of ignore IA:\  we only used the observed $w^{\gamma\kappa}$, while  including $A_{\rm lens}$ in the fit and ignoring the contamination by IA (by setting $A_{\rm IA}=0$); (2)  case of IA w/o SC:\ we only used the observed $w^{\gamma\kappa}$, but consider both $A_{\rm lens}$ and $A_{\rm IA}$ following Eq.\,\ref{eq w^GK fit}; (3) case of SC:\ we used both $w^{\gamma\kappa}$ in Fig.\,\ref{fig GK} and the SC correlations in Fig.\,\ref{fig GgIg}. Both the CMB lensing amplitude $A_{\rm lens}$ and the nuisance parameters \{$A_{\rm IA}$, $b_{\rm g,eff}$, $g_{\rm mag}$\} are used in the analysis, following Eqs.\,\ref{eq w^Gg fit}, \ref{eq w^Ig fit}, and \ref{eq w^GK fit}.

\begin{figure}\centering
        \includegraphics[width=1.0\columnwidth]{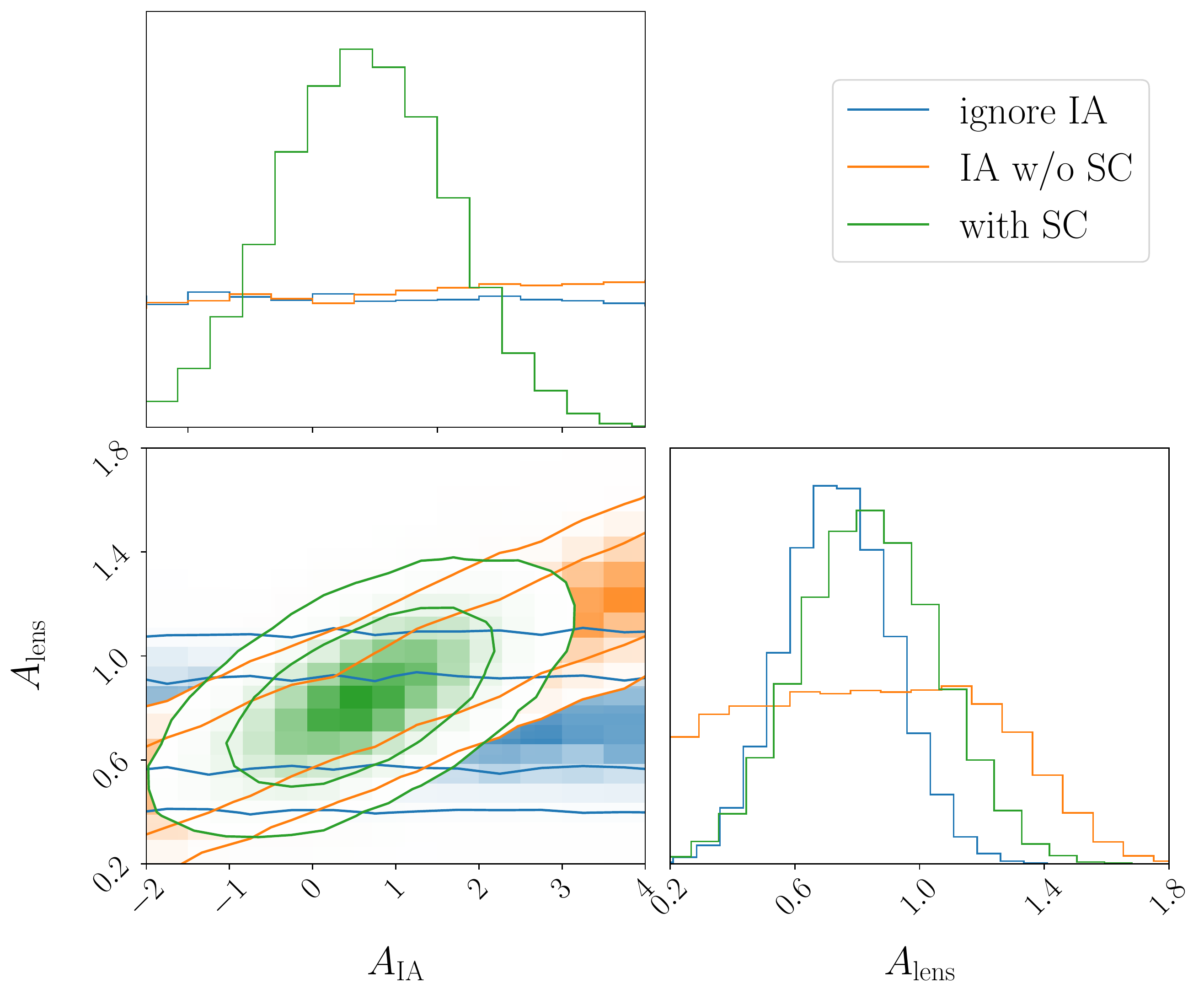}
        \caption{Constraints on lensing amplitude $A_{\rm lens}$ and the IA amplitude $A_{\rm IA}$, with three different methods: assuming there is no IA in the measured $w^{\kappa\gamma}$ (blue), considering the impact of IA with conventional IA model but do not use SC (orange), and using SC to subtract IA information and constrain together with the CMB lensing cross-correlation (green). When IA is ignored, $A_{\rm IA}$ is unconstrained. The similar height and width of $A_{\rm lens}$ PDFs between blue and green prove that by including SC, the $A_{\rm IA}-A_{\rm lens}$ degeneracy can be efficiently broken so that the constraining power loss in $A_{\rm lens}$ is very small.}
        \label{fig MCMC}
\end{figure}

\begin{figure}\centering
        \includegraphics[width=1.0\columnwidth]{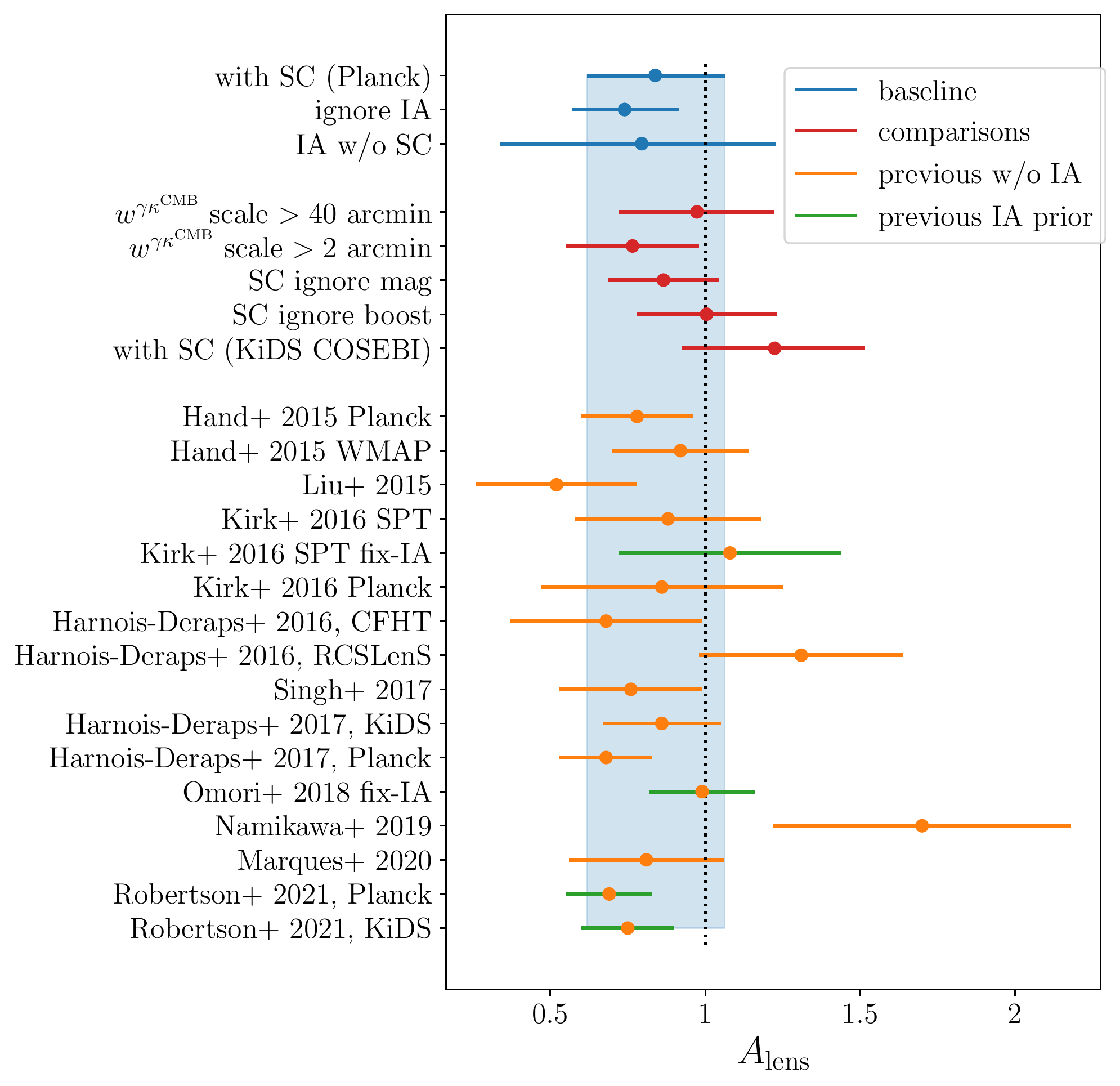}
        \caption{Comparisons of the constraints on $A_{\rm lens}$ with previous measurements. Our baseline analysis ``with SC'' is consistent with 1. We also show some cases where IA is ignored in the analysis and if IA is considered but the $A_{\rm IA}-A_{\rm lens}$ degeneracy is not broken with SC. These main results in blue are similar to what is shown in Fig.\,\ref{fig MCMC}. We show tests with different scale cuts and different treatments to magnification, the boost factor, and different (KiDS) fiducial cosmologies in red. We make a comparison with other works, divided into that ignoring IA (orange) and that assuming a strong prior of IA (green). We note that for different works, the different fiducial cosmologies (the ``Planck,'' ``WMAP,'' and ``KiDS'' labels on the y-axis) can lead to $\sim10\%$ difference in $A_{\rm lens}$.}
        \label{fig compareA}
\end{figure}

\begin{figure}\centering
        \includegraphics[width=1.0\columnwidth]{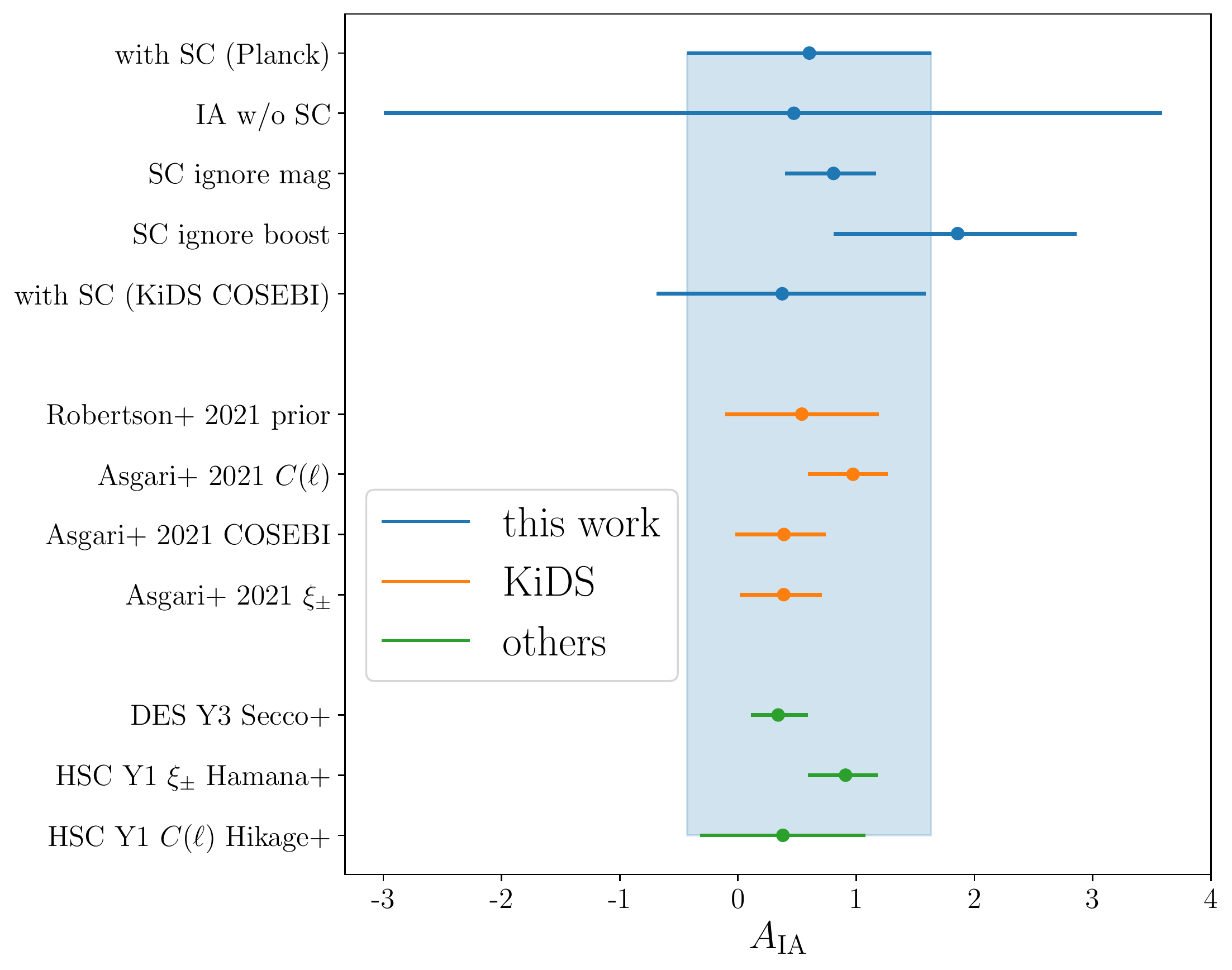}
        \caption{Comparisons of the constraints on $A_{\rm IA}$. We show the results of this work in blue, which contains our fiducial analysis with SC applied and the comparisons of (1) without SC, (2) with SC but ignoring magnification, (3) with SC but ignoring the boost factor, and (4) switching to the KiDS fiducial cosmology. We show comparisons with other works using KiDS-1000 data in orange and some works using DES or HSC data in green.}
        \label{fig compareAIA}
\end{figure}

The results of these simulations are shown in Fig.\,\ref{fig MCMC}. We use flat priors in $0<A_{\rm lens}<2$, $-5<A_{\rm IA}<5$, and for the IA self-calibration nuisance parameters we use $0<b_{\rm g,eff}<4$, $-5<g_{\rm mag}<5$. For case 1 (ignore IA), shown in blue, $A_{\rm IA}$ is unconstrained in the fitting, giving the best-fit $A_{\rm lens}=0.74^{+0.18}_{-0.17}$.  For case 2 (IA w/o SC, when we consider the existence of IA and apply the IA model as in Eq.\,\ref{eq IA P(k) model}, but do not use the measurements from SC (Fig.\,\ref{fig GgIg} and Eq.\,\ref{eq w^Gg fit}, \ref{eq w^Ig fit}), there will be a strong degeneracy between $A_{\rm lens}$ and $A_{\rm IA}$, as shown in orange. There is a significant loss of constraining power in the lensing amplitude, with the best-fit $A_{\rm lens}=0.79^{+0.43}_{-0.46}$ and $A_{\rm IA}=0.47^{+3.11}_{-3.47}$. For case 3 (with SC), the introduced measurements of $w^{\rm Gg}$ and $w^{\rm Ig}$ has the capacity to not only break the degeneracy between $A_{\rm lens}$ and $A_{\rm IA}$ (see Eqs.\,\ref{eq w^Gg fit}, \ref{eq w^Ig fit}, and \ref{eq w^GK fit}), but also to bring more constraining power to $A_{\rm IA}$, so that the best fit of $A_{\rm lens}$ will  be both unbiased (according to the validation using simulation) and have significantly improved constraining power. The best-fit values are $A_{\rm lens}=0.84^{+0.22}_{-0.22}$, $A_{\rm IA}=0.60^{+1.03}_{-1.03}$, $b_{\rm g,eff}=0.88^{+0.06}_{-0.06}$, and $g_{\rm mag}=-0.30^{+1.60}_{-1.62}$. In Fig.\,\ref{fig MCMC}, we only show $A_{\rm IA}$ and $A_{\rm lens}$, which are the focus of this work, while $b_{\rm g,eff}$ and $g_{\rm mag}$ are only related with the SC observables but not CMB lensing. Also, as discussed in \cite{Yao2020}, the existence of the effective galaxy bias $b_{\rm g,eff}$ can also absorb some systematics (thus, it may be a biased bias), leaving the constraint on $A_{\rm IA}$ unbiased (as shown in Fig.\,\ref{fig compareAIA_MICE}). For example, we tested whether when magnification is absent, the effect of the boost factor will be fully absorbed by $b_{\rm g,eff}$, giving unbiased values for $A_{\rm IA}$ and $A_{\rm lens}$. The effective galaxy bias could also absorb the differences in the assumed fiducial cosmology, with $b_{\rm g,eff}\sim1.24$ with KiDS COSEBI cosmology, for example. The redshift distribution $n(z)$ can differ slightly when accounting for (or not) the lensing weight (considering the lensing or clustering part in the galaxy-shape correlation), with a $\sim0.024$ difference in the mean-z, which can lead to $\sim8\%$ difference in the theoretical lensing signal and $\sim2\%$ difference in the theoretical IA signal. Other unaddressed sources of systematics such as baryonic feedback and massive neutrinos could have similar effects. We can also see from the validation using MICE data that although the resulting $b_{\rm g,eff}$ is lower than the expectation, the $A_{\rm IA}$ result is unbiased. The $g_{\rm mag}$ result also resides in a reasonable range, considering the KiDS i-band magnitude \citep{Kuijken2019} and comparing it with \cite{Duncan2014}. The above three cases of IA treatments are also summarized later in Figs.\,\ref{fig compareA} and \ref{fig compareAIA} together with more tests and other works.

The corresponding best-fit curves are shown in Figs.\,\ref{fig Cell} and \ref{fig GgIg} with $A_{\rm IA}=0.60^{+1.03}_{-1.03}$, $b_{\rm g,eff}=0.88^{+0.06}_{-0.06}$, and $g_{\rm mag}=-0.30^{+1.60}_{-1.62}$ . Even though the impact of magnification is comparable to the IA signal, we can see in both the angular power spectrum and correlation function that the shapes of IA and magnification are different. For example, as shown in Fig.\,\ref{fig GgIg}, the tidal alignment model $w^{\rm Ig}$ and magnification $g_{\rm mag}w^{\rm G\kappa}$ are comparable at large scales, but differ at small scales. Therefore, in principle, the degeneracy between IA and magnification can be broken for future data with higher S/N, so that the shape or slope information of the observables can be used. The current degeneracy is due to the low S/N so that the amplitudes of $A_{\rm IA}$ and $g_{\rm mag}$ degenerate. Furthermore, if a more complicated IA model is used, for example, as in \cite{Blazek2017,DESY3cosmo}, the small-scale IA will be different. Based on the study of \cite{Shi2021}, for a wide range of stellar mass, the small-scale IA should have a higher amplitude (either a direct raise in the amplitude or a ``drop-raise'' pattern as we go down to smaller scales) than the current model so that the IA-magnification degeneracy can be broken further. The appropriate IA model will require studies in many aspects and with higher S/N in the measurements. Thus, we leave this topic to a future work.

We investigate how different choices can change our results. We first compare the different scale cuts for $w^{\kappa\gamma}$. Besides the baseline analysis of $A_{\rm lens}=0.84^{+0.22}_{-0.22}$ with $\theta>20$ arcmin, two more tests are made with a larger scale cut of $\theta>40$ arcmin and a smaller scale cut of $\theta>2$ arcmin, as shown in Fig.\,\ref{fig GK}, which give us $A_{\rm lens}=0.97^{+0.25}_{-0.25}$ and $A_{\rm lens}=0.77^{+0.21}_{-0.22}$, respectively. The comparisons are shown in Fig.\,\ref{fig compareA}. The large-scale lensing amplitude is higher than the small-scale one, which agrees with the finding in \cite{Planck2018lensing} and other works on cross-correlations \citep{Sun2022}. In this work, we only report this large-scale vesus small-scale difference. However, the current S/N of the correlation of the CMB convergence and galaxy shear and the model assumptions do not allow us to investigate further on this topic. Similarly, the impact from point-source subtraction errors in the CMB lensing map \citep{Planck2018lensing} is negligible ($\le0.2\%$) in this cross-correlation analysis.

We then compare the different choices in the SC method. We find that if the magnification model is ignored in the analysis, the existing magnification signal in the data will be treated as an IA signal, leading to an overestimated $A_{\rm IA}=0.81^{+0.36}_{-0.41}$ and an overestimated $A_{\rm lens}=0.87^{+0.18}_{-0.18}$. On the other hand, we previously argued that when magnification is absent, the impact from the boost factor will be purely absorbed by the effective galaxy bias, $b_{\rm g,eff}$, leaving $A_{\rm IA}$ and $A_{\rm lens}$ unbiased. Unfortunately, this does not hold anymore when magnification is present: if the boost factor is not corrected, all the parameters will be biased as follows $A_{\rm IA}=1.86^{+1.01}_{-1.05}$, $b_{\rm g,eff}=0.67^{+0.06}_{-0.06}$, $A_{\rm lens}=1.00^{+0.23}_{-0.23}$ and $g_{\rm mag}=1.55^{+1.28}_{-1.31}$. We include the comparisons of $A_{\rm lens}$ and $A_{\rm IA}$ for the above-described cases in Fig.\,\ref{fig compareA} and \ref{fig compareAIA} and emphasis the importance of taking magnification and boost factor into consideration. We also show the impact of the assumed fiducial cosmology: if the fiducial cosmology is switched from {\it Planck} to KiDS-1000 COSEBI as in Table \ref{table fiducial cosmology}, both $A_{\rm lens}$ and $A_{\rm IA}$ will change as shown in Fig.\,\ref{fig compareA} (bottom:\ red) and \ref{fig compareAIA} (bottom:\ blue).

\begin{figure}\centering
        \includegraphics[width=1.0\columnwidth]{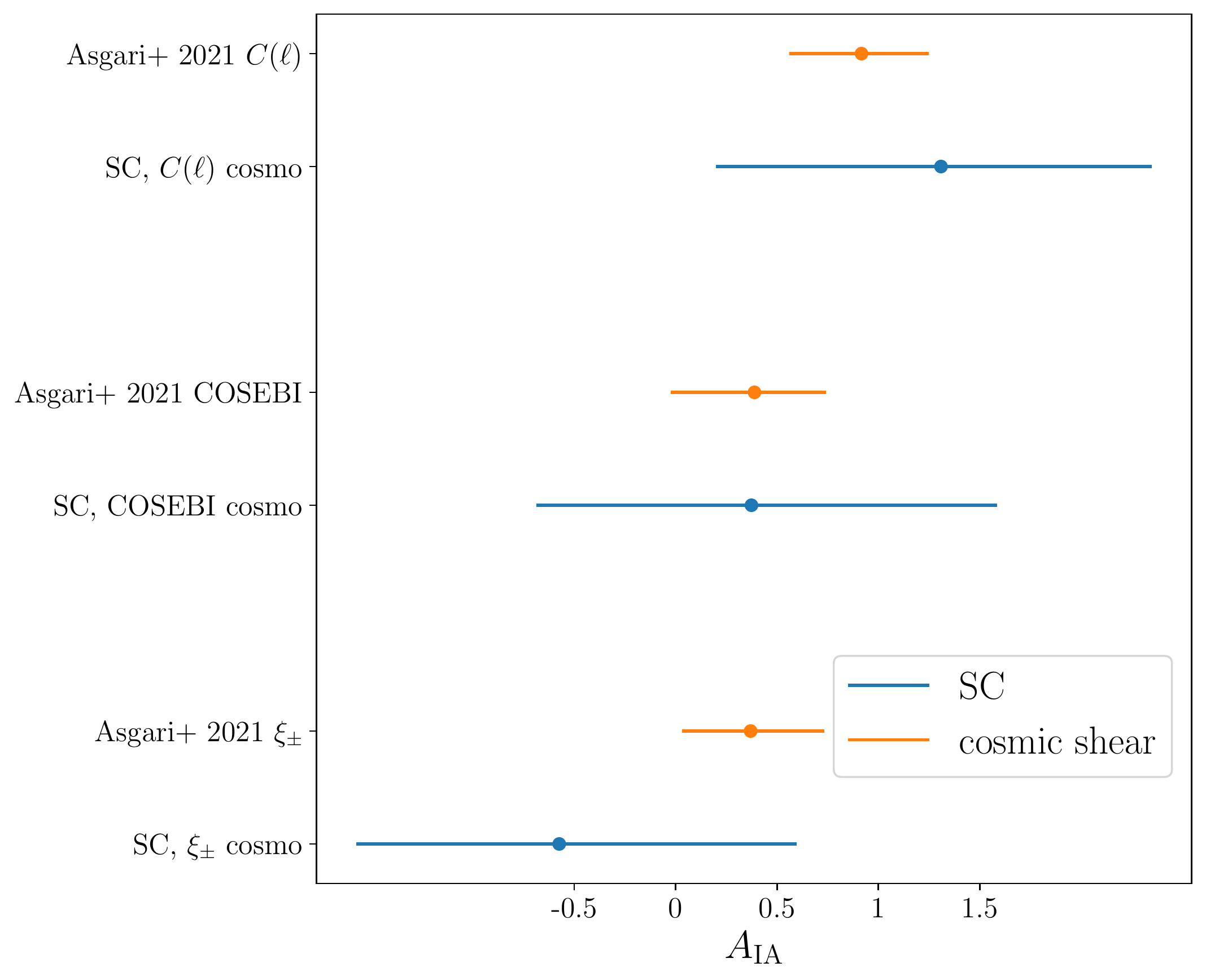}
        \caption{Comparisons of $A_{\rm IA}$ between SC-subtracted results (blue) and cosmic shear tomography subtracted results (orange) with cosmologies from different two-point statistics. The cosmologies are shown in Table \ref{table fiducial cosmology}.}
        \label{fig compareAIAcosmo}
\end{figure}

With the above results in simulation and data, summarized in Figs.\,\ref{fig compareAIA_MICE}, \ref{fig compareA}, and \ref{fig compareAIA}, we show that our measurements on $A_{\rm IA}$ and $A_{\rm lens}$ are unbiased with regard to magnification, the boost factor, and the assumed photo-z PDF model. These are new developments that consider the existence of magnification at high redshift $z\sim1$, beyond the study of \cite{Yao2020}.

Additionally, we compare our analysis with previous works. The comparisons of $A_{\rm lens}$ are shown in Fig.\,\ref{fig compareA}. We find that most of the previous works have ignored the IA contamination \citep{Hand2015,Liu2015,Kirk2016,Harnois-Deraps2016,Singh2017b,Harnois-Deraps2017,Namikawa2019,Marques2020}. For the ones that considered IA, they either fixed the IA amplitude \citep{Kirk2016,Omori2019} or used a strong prior \citep{Robertson2021} to break the degeneracy between $A_{\rm lens}$ and $A_{\rm IA}$, which would otherwise cause a strong loss in constraining power, as we show in Fig.\,\ref{fig MCMC}. We are the first to directly achieve the IA amplitude measurement within the same data and break the lensing-IA degeneracy. Our baseline analysis is consistent with most of the previous results, showing the contamination from IA is not significant, mainly due to the total S/N of CMB lensing-galaxy shear cross-correlation is only at $3\sim5$ $\sigma$ level at the current stage. However, the correct treatment for IA will be more and more important in the future with the advent of stage IV cosmic shear surveys and CMB observations.

Comparisons of the $A_{\rm IA}$ constraint with other results using KiDS-1000 data are shown in Fig.\,\ref{fig compareAIA}, including the prior assumed in \cite{Robertson2021} and the cosmic shear tomography constraint in \cite{Asgari2021}. Although the redshift range is slightly different, the above works demonstrate consistent results for $A_{\rm IA}$. These comparisons will become more interesting for the next-stage observations.

As an extended study, we investigate how the choice of fiducial cosmology affects the SC results, namely $A_{\rm IA}$. In Fig.\,\ref{fig compareAIA}, we show the results with the fiducial {\it Planck} cosmology and the KiDS-1000 two-point correlation function $\xi_\pm$ best-fit cosmology. We further compare the results with the KiDS-1000 band power $C(\ell)$ cosmology and the COSEBIs cosmology in Fig.\,\ref{fig compareAIAcosmo}. The results from \cite{Asgari2021} (shown in orange) are arranged in increasing order from bottom to top. We find that when assuming the same cosmology, the SC results (shown in blue) also follow the same (weak) trend; meanwhile, they agree very well with the cosmic shear results. We note the SC results will provide extra information in constraining IA in cosmic shear in the future. 

\section{Summary} \label{Section summary}

In this work, we carry out the first application of the self-calibration (SC) method of intrinsic alignment (IA) of galaxies to its cosmological application. We proved that with SC, the lensing-IA degeneracy could be efficiently broken, namely, in this CMB lensing $\times$ galaxy shear cross-correlation work, it means breaking the degeneracy between the lensing amplitude, $A_{\rm lens}$, and the IA amplitude, $A_{\rm IA}$. We show that for previous treatments, IA values are either ignored or being considered with a strong assumed prior on $A_{\rm IA}$. We demonstrate in Figs.\,\ref{fig MCMC}, \ref{fig compareA}, and \ref{fig compareAIA} that with SC to break the degeneracy, the constraining power in both $A_{\rm lens}$ and $A_{\rm IA}$ is preserved.

We demonstrate that the proper angular scale cuts on $w^{\kappa\gamma}$ are important. Our baseline analysis using information from $\theta>20$ arcmin gives $A_{\rm lens}=0.84^{+0.22}_{-0.22}$. If we use information only at larger scales with $\theta>40$ arcmin, the constraint is $A_{\rm lens}=0.97^{+0.25}_{-0.25}$. If we include information at much smaller scales with $\theta>2$ arcmin, the constraint is $A_{\rm lens}=0.77^{+0.21}_{-0.22}$. At the current stage, these results do not differ significantly from each other (even considering they are strongly correlated), as shown in Fig.\,\ref{fig compareA}. However, we note that these differences at different scales also exist in other works, such as \cite{Planck2018lensing} and \cite{Sun2022}. Therefore, we emphasize the importance of understanding the possible systematics at different scales for future studies with higher S/N, possibly using updated maps with more data \citep{Carron2022} or in combination with other observations \citep{Robertson2021,Darwish2021}.

Comparing our CMB lensing amplitude, $A_{\rm lens}$, with other works in Fig.\,\ref{fig compareA}, we found consistent results with different treatments of IA throughout almost all the works. We conclude that IA is not a significant source of systematics for the current stage. However, it will soon become more important with the stage IV observations. Nevertheless, we emphasize that the correct treatment to break the lensing-IA degeneracy is very important for maintaining the cosmological constraining power. Our constraint on the IA amplitude $A_{\rm IA}$ in Fig.\,\ref{fig compareAIA} is also consistent with the existing analysis on IA with KiDS-1000 data. We note that the SC-subtracted IA information can be used as extra constraining power for any of these analyses. 

On the technique side, we further developed the SC method considering more sources of systematics beyond \cite{Yao2020}. We show that at $z\sim1$, the impact of galaxy shear $\times$ cosmic magnification component $w^{\rm G\kappa^{\rm gal}}$ contaminates the separated IA $\times$ galaxy number density signal $w^{\rm Ig}$ and is non-negligible, as shown in Figs.\,\ref{fig Cell} and \ref{fig GgIg}. We use Eq.\,\ref{eq Ig correlation} and \ref{eq w^Ig fit} to show how the magnification term enters our observable and how we include it in the theory as a correction. We show in Figs.\,\ref{fig compareA} and \ref{fig compareAIA} that the correction of magnification is important when applying SC to higher redshift data in order to get the correct constraint on IA. We also discuss the fact that with the contamination from magnification, the boost factor can no longer be absorbed by the effective galaxy bias $b_{\rm g,eff}$, and need to be accounted for correctly, as shown in Eqs.\,\ref{eq gamma-g estimator} and \ref{eq boost} and Fig.\,\ref{fig GgIg}, \ref{fig compareA}, and \ref{fig compareAIA}.

We also validated our analysis with MICE2 simulation, focusing on two aspects: (1) how well  magnification model can mitigate the contamination from the magnification-shear signal and (2) whether the assumed photo-z PDF model (which is used to calculate the signal drop $Q^{\rm Gg}$ and $Q^{\rm Ig}$) bias the IA measurement. With the strong constraining power from MICE2 with no shape noise, we can show in Fig.\,\ref{fig compareAIA_MICE} that, when the magnification model is included in the analysis, the IA amplitude can be obtained correctly (consistent within $1\sigma$ range of 0, which is the input of MICE2). Additionally, the bias from the assumed photo-z model is negligible when the magnification model is used, as the effective magnification prefactor $g_{\rm mag}$ will absorb the introduced error, leaving the residual bias $|\Delta A_{\rm IA}|<0.1$. This validation is suitable for KiDS and KiDS-like data only. Therefore, we emphasize the importance of including the magnification model and advanced photo-z modeling in the SC analysis, especially for future high-z surveys such as LSST, Euclid, WFIRST, and CSST. We further notice the contamination from magnification will make SC no longer an IA-model-independent method, therefore, SC is more suitable for low-z data when considering alternative IA models.

In comparison with our first measurements with the KV-450 data \citep{Yao2019}, a lot of improvements have been made to the SC method, including: (1) the covariance, the galaxy bias, the scale-dependency for the lensing-drop $Q^{\rm Gg}$, the IA-drop $Q^{\rm Ig}$, and appropriate scale-cuts, which have been introduced in \cite{Yao2020}; (2) the boost factor, the cosmic magnification, and the photo-z PDF modeling, which are introduced in this work; (3) its first validation using simulation, and its first application to cosmology in order to break the lensing-IA degeneracy, introduced in this work. With these improvements, we manage to achieve consistent IA results between SC and cosmic shear, as shown in Fig.\,\ref{fig compareAIAcosmo}, while previously we got $A_{\rm IA}=2.31^{+0.42}_{-0.42}$ with the old version of SC \citep{Yao2019} and $A_{\rm IA}=0.981^{+0.694}_{-0.678}$ for cosmic shear \citep{Hildebrandt2020} with KV-450 data.

The SC-obtained $A_{\rm IA}$ is consistent with the MICE-input IA as well as with the KiDS
cosmic shear results when applied to the data, for instance, from \cite{Asgari2021}.  It is also consistent with the other CMB lensing works, such as \cite{Robertson2021},  and $g_{\rm mag}$ is also in reasonable agreement with the work of \cite{Duncan2014}. Nevertheless, our results still suffer from an unrealistically low effective galaxy bias of $b_{\rm g,eff}=0.88$, which is different from the results presented in our previous work \citep{Yao2020}. We discuss the possibility that this value may absorb the contribution from (1) fiducial cosmology, (2) lensing weight in $n(z)$, (3) insufficient modeling in non-linear galaxy bias, baryonic effects, and massive neutrinos, (4) incorrect photo-z versus true-z connection (discussed in Appendix\,\ref{Appendix Q}), and (5) possible other sources of systematics. We emphasize these complications and leave this point for further discussion in future studies. 

We note that other systematics may be at work other than the galaxy bias, such as the beyond Limber approximation \citep{Fang2020}, non-flat $\Lambda$CDM \citep{Yu2021}, or selection bias on shear measurements \citep{Li2021}. However, these have either a much smaller impact compared to IA or they are strongly reduced due to our scale cuts. Thus, these remain beyond the scope of this paper and may be the topic of a future work.

\begin{acknowledgements}
        
        The authors thank Yu Yu, Hai Yu, Jiaxin Wang for useful discussions. \\
        
        This work is supported by National Key R\&D Program of China No. 2022YFF0503403. JY, HYS, PZ and XL acknowledge the support from CMS-CSST-2021-A01, CMS-CSST-2021-A02 and CMS-CSST-2021-B01.
        JY acknowledges the support of the National Science Foundation of China (12203084), the China Postdoctoral Science Foundation (2021T140451), and the Shanghai Post-doctoral Excellence Program (2021419). 
        HYS acknowledges the support from NSFC of China under grant 11973070, the Shanghai Committee of Science and Technology grant No.19ZR1466600 and Key Research Program of Frontier Sciences, CAS, Grant No. ZDBS-LY-7013. 
        PZ acknowledges the support of the National Science Foundation of China (11621303, 11433001). 
        XL acknowledges the support of NSFC of China under Grant No. 12173033, 11933002, 11803028, YNU Grant No. C176220100008, and a grant from the CAS Interdisciplinary Innovation Team. 
        BJ acknowledges support by STFC Consolidated Grant ST/V000780/1. 
        MB is supported by the Polish National Science Center through grants no. 2020/38/E/ST9/00395, 2018/30/E/ST9/00698, 2018/31/G/ST9/03388 and 2020/39/B/ST9/03494, and by the Polish Ministry of Science and Higher Education through grant DIR/WK/2018/12. 
        HH is supported by a Heisenberg grant of the Deutsche Forschungsgemeinschaft (Hi 1495/5-1) as well as an ERC Consolidator Grant (No. 770935). 
        TT acknowledges support from the Leverhulme Trust. 
        AW is supported by an European Research Council Consolidator Grant (No. 770935). 
        ZY acknowledges support from the Max Planck Society and the Alexander von Humboldt Foundation in the framework of the Max Planck-Humboldt Research Award endowed by the Federal Ministry of Education and Research (Germany).
        The computations in this paper were run on the $\pi$ 2.0 cluster supported by the Center for High Performance Computing at Shanghai Jiao Tong University.\\
        
        The codes JY produced for this paper were written in Python. JY thanks all its developers and especially the people behind the following packages: SCIPY \citep{scipy}, NUMPY \citep{numpy}, ASTROPY \citep{astropy} and MATPLOTLIB \citep{matplotlib}, TreeCorr \citep{Jarvis2004}, CCL \citep{Chisari2019CCL}, CAMB \citep{Lewis2000CAMB}, Healpy \citep{Healpy_Gorski2005,Healpy_Zonca2019}, emcee \citep{emcee}, fitsio\footnote{\url{https://github.com/esheldon/fitsio}}, kmeans\_radec\footnote{\url{https://github.com/esheldon/kmeans_radec}}, corner \citep{corner}, ChainConsumer\footnote{\url{https://github.com/Samreay/ChainConsumer}}.\\
        
        The KiDS-1000 results in this paper are based on data products from observations made with ESO Telescopes at the La Silla Paranal Observatory under programme IDs 177.A-3016, 177.A-3017 and 177.A-3018, and on data products produced by Target/OmegaCEN, INAF-OACN, INAF-OAPD, and the KiDS production team, on behalf of the KiDS consortium.\\
        
        Author contributions: All authors contributed to the development and writing of this paper. The authorship list is given in three groups: the lead authors (JY, HS, PZ, XL) followed by two alphabetical groups. The first alphabetical group includes those who are key contributors to both the scientific analysis and the data products. The second group covers those who have either made a significant contribution to the data products, or to the scientific analysis.\\
\end{acknowledgements}

%
\bibliographystyle{aa} 
\bibliography{references} 
%

\begin{appendix} 
        \section{Signal drop Q} \label{Appendix Q}
        
        We keep the main text of this paper focused on the physics and presenting the details of the SC method, more specifically the calculation for the lensing-drop $Q^{\rm Gg}$ and the IA-drop $Q^{\rm Ig}$, in this appendix. The $Q$s are calculated through Eq.\,\ref{eq Q^Gg theta} and \ref{eq Q^Ig theta}, while the correlation functions being used are just the Hankel transform (similar to Eq.\,\ref{eq w Hankel}) of the angular power spectrum $C^{\rm Gg}$ and $C^{\rm Ig}$. The associated $C^{\rm Gg}$ and $C^{\rm Gg}|_{\rm S}$ are calculated via:
        \begin{align}
                C^{\rm Gg}_{ii}(\ell) &= \int_0^\infty\frac{q_i(\chi)n_i(\chi)}{\chi^2} b_{\rm g,eff} P_\delta\left(k=\frac{\ell}{\chi};\chi\right)d\chi, \label{eq C^Gg} \\
                C^{\rm Gg}_{ii}|_{\rm S}(\ell) &= \int_0^\infty\frac{q_i(\chi)n_i(\chi)}{\chi^2}
                b_{\rm g,eff} P_\delta\left(k=\frac{\ell}{\chi};\chi\right) \eta^{\rm Gg}_i(z) d\chi. \label{eq C^Gg_S}
        \end{align}
        Similarly, the $C^{\rm Ig}$ and $C^{\rm Ig}|_{\rm S}$ are given by:
        \begin{align}
                C^{\rm Ig}_{ii}(\ell) &= \int_0^\infty\frac{n_i(\chi)n_i(\chi)}{\chi^2}b_{\rm g,eff}P_{\delta,\gamma^I}\left(k=\frac{\ell}{\chi};\chi\right)d\chi, \label{eq C^Ig} \\
                C^{\rm Ig}_{ii}|_{\rm S}(\ell) &= \int_0^\infty\frac{n_i(\chi)n_i(\chi)}{\chi^2}b_{\rm g,eff}P_{\delta,\gamma^I}\left(k=\frac{\ell}{\chi};\chi\right)\eta^{\rm Ig}_i(z) d\chi. \label{eq C^Ig_S}
        \end{align}
        Here, $\eta^{\rm Gg}_i(z)=\eta^{\rm Gg}_i(z_{\rm L}=z_g=z)$ is the function that account for the effect of the SC selection Eq.\,\ref{eq SC selection} in the Limber integral, similarly for $\eta^{\rm Ig}$. They are expressed as:\ 
                \begin{align}
                        \eta^{\rm Gg}_i(z_{\rm L},z_g) &= \frac 
                        {2\int dz^{\rm P}_{G}\int dz^{\rm P}_{\rm g}\int_{0}^{\infty}dz_{\rm G} W_L(z_{\rm L},z_{\rm G})S(z^{\rm P}_{\rm G},z^{\rm P}_{\rm g})K}
                        {\int dz^{\rm P}_{G}\int dz^{\rm P}_{\rm g}\int_{0}^{\infty}dz_{\rm G}
                                W_L(z_{\rm L},z_{\rm G})K} ,
                        \label{eq eta^Gg} \\
                        \eta^{\rm Ig}_i(z_{\rm L},z_g) &= \frac 
                        {2\int dz^{\rm P}_{G}\int dz^{\rm P}_{\rm g}\int_{0}^{\infty}dz_{\rm G} S(z^{\rm P}_{\rm G},z^{\rm P}_{\rm g})K}
                        {\int dz^{\rm P}_{G}\int dz^{\rm P}_{\rm g}\int_{0}^{\infty}dz_{\rm G}
                                K} , \label{eq eta^Ig}
                \end{align}
        as in \cite{Yao2020}, where $K$ is the galaxy-pair redshift distribution kernel:        \begin{equation}
                K(z_{\rm G},z_g,z^{\rm P}_{\rm G},z^{\rm P}_{\rm g})=p(z_{\rm G}|z^{\rm P}_{\rm G})p(z_g|z^{\rm P}_{\rm g})n^{\rm P}_i(z^{\rm P}_{\rm G})n^{\rm P}_i(z^{\rm P}_{\rm g}), \label{eq K kernel}
        \end{equation}
        and $S$ is the SC selection function:
        \begin{equation} \label{eq SC selection func}
                S(z^{\rm P}_{\rm G},z^{\rm P}_{\rm g})=\begin{cases}
                        1 &\text{for $z^{\rm P}_{\rm G}<z^{\rm P}_{\rm g}$},\\
                        0 &\text{otherwise}\ ,
                \end{cases}
        \end{equation}
        which correspond to Eq.\,\ref{eq SC selection} in the main text, and the lensing kernel is: 
        \begin{equation}
                W_L(z_{\rm L},z_S)=\begin{cases}
                        \frac{3}{2}\Omega_{\rm m}\frac{H_0^2}{c^2}(1+z_{\rm L})\chi_{\rm L}(1-\frac{\chi_{\rm L}}{\chi_{\rm S}}) &\text{for $z_{\rm L}<z_S$}\\
                        0 &\text{otherwise}
                \end{cases}.
        \end{equation}
        Here, $z_x$ is the true-z where $x$ can be ``G'' the source, ``L'' the lens, or ``g'' the galaxy number density. The galaxy photo-z distribution is $n^{\rm P}(z^{\rm P})$, and the redshift PDF (probability distribution function) is $p(z|z^{\rm P})$.
        
        As shown above, when the galaxy photo-z distribution and the corresponding true-z distribution are given, as shown in Fig.\,\ref{fig nz} in this work, we can follow the above procedure to calculate the lensing-drop $Q^{\rm Gg}$ and $Q^{\rm Ig}$. The results of $Q^{\rm Gg}$ and $Q^{\rm Ig}$ for this work are shown in Fig.\,\ref{fig Q}. Generally, given the tomographic bin width, the better photo-z is, the smaller $Q^{\rm Gg}$ will be (it reaches $\sim0$ for perfect photo-z). On the other hand, non-symmetric photo-z distribution and non-symmetric true-z distribution will make $G^{Ig}$ deviate from 1. For more details on the $Q$ calculation and its properties, we refer to the discussions in \cite{Yao2019,Yao2020}.
        
        We note that for the SC calculation, the redshift PDF $p(z|z^{\rm P})$ for each galaxy is required. Due to the fact that the PDFs from photo-z algorithm can be biased due to the color-redshift degeneracy in the photometric surveys, calibration is needed \citep{Hildebrandt2016,Hildebrandt2021,DESY3cosmo}. However, we can only statistically calibrate the overall redshift distribution $n(z)$ but not the PDF $p(z|z^{\rm P})$ for each galaxy. This means in order to calculate Eq.\,\ref{eq K kernel}, we need to assume a photo-z PDF model. We chose to use a bi-Gaussian model \cite{Yao2019}
        \begin{equation}
                p_{2G}(z|z^{\rm P})=(1-f_{\rm out})p_{\rm main}(z|z^{\rm P};\Delta_1,\sigma_1)+f_{\rm out}p_{\rm outlier}(z|z^{\rm P};\Delta_2,\sigma_2) , \label{eq bi-Gaussian}
        \end{equation}
        with a main Gaussian peak and a Gaussian outlier peak with different bias, $\Delta_i$, and scatter, $\sigma_i$, and an outlier rate, $f_{\rm out}$.
        
        \begin{figure}\centering
                \includegraphics[width=1.0\columnwidth]{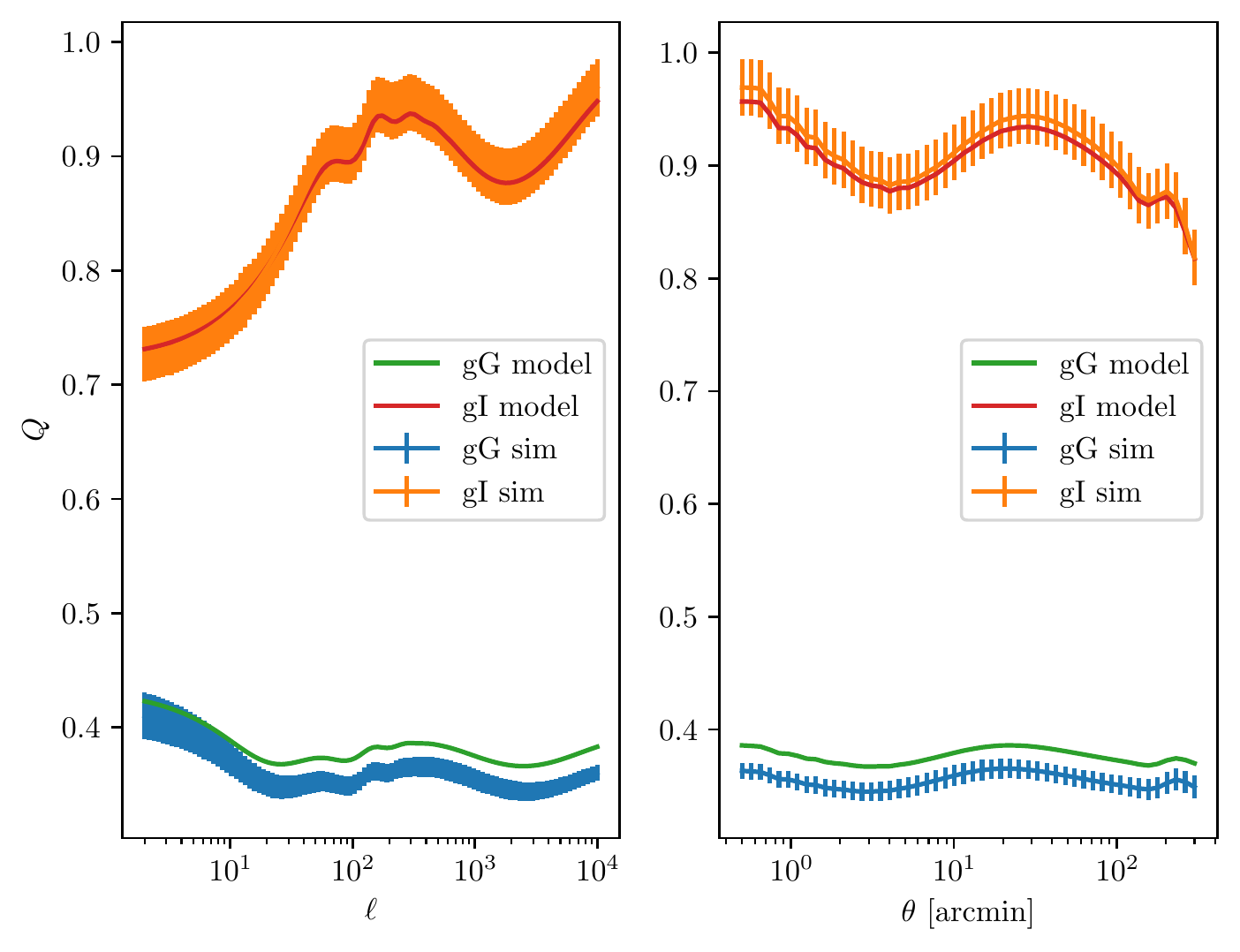}
                \caption{Effect of photo-z modeling with MICE2.  By applying the SC selection in Eq.\,\ref{eq SC selection} or \ref{eq SC selection func}, the lensing-drop $G^{Gg}$ from photo-z model (green) is slightly biased compared with the results from true-z (blue), while the IA-drop $G^{Ig}$ from photo-z model (red) is immune to such bias and agrees with the true-z result (orange).}
                \label{fig Q MICE}
        \end{figure}
        
        We fit the bi-Gaussian model Eq.\,\ref{eq bi-Gaussian}, requiring it to have same mean redshift $\left<z\right>$ with the SOM calibrated $n(z)$ \citep{Asgari2021}, and minimized the difference between the resulting model z-distribution $\int n^{\rm P}(z^{\rm P})p(z|z^{\rm P})dz^{\rm P}$ and the SOM $n(z)$. The best fit will then be a good description of the photo-z quality and can be used in Eq.\,\ref{eq K kernel}. The resulting signal drops are shown in Fig.\,\ref{fig Q} in the main text.
        
        We validated the bi-Gaussian photo-z model for SC with MICE2 simulation. We made a comparison with the results that use the photo-z distribution and true-z distribution in the calculation of Eq.\,\ref{eq K kernel}. We show in Fig.\,\ref{fig Q MICE} that the bi-Gaussian model can produce the IA-drop $Q^{\rm Ig}$ measurement very consistently with those using true-z from simulations. However, we find the lensing-drop $Q^{\rm Gg}$ from the photo-z model is slightly higher than the true values from the simulation. This error will be propagated to the separated lensing signal $w^{\rm Gg}$ and the IA+magnification signal $w^{\rm Ig}+g_{\rm mag}w^{\rm G\kappa}$, according to Eqs.\,\ref{eq Gg correlation} and \ref{eq Ig correlation}. Its impact in $A_{\rm IA}$ is shown in Figs.\,\ref{fig MICE Qmodel} and \ref{fig compareAIA_MICE}.
        
\end{appendix}

\end{document}